\documentclass{iopart}

\usepackage{graphicx}
\usepackage{mathrsfs}
\usepackage{hyperref}
\usepackage{geometry}
\newcommand{\text}[1]{\mathrm{#1}}

\begin{document}

\title[Spontaneous Skyrmion Conformal Lattice...]{Spontaneous Skyrmion Conformal Lattice and Transverse Motion During dc and ac Compression}
\author{J.~C.~Bellizotti Souza$^1$, 
            N.~P.~Vizarim$^{1,2}$, 
            C.~J.~O.~Reichhardt$^3$, 
            C.~Reichhardt$^3$
            and P.~A.~Venegas$^4$}
\address{$^1$ POSMAT - Programa de P\'os-Gradua\c{c}\~ao em Ci\^encia e Tecnologia de Materiais, Faculdade de Ci\^encias, Universidade Estadual Paulista - UNESP, Bauru, SP, CP 473, 17033-360, Brazil}
\address{$^2$ Department of Physics, University of Antwerp, Groenenborgerlaan 171, B-2020 Antwerp, Belgium}
\address{$^3$ Theoretical Division and Center for Nonlinear Studies, Los Alamos National Laboratory, Los Alamos, New Mexico 87545, USA}
\address{$^4$ Departamento de F\'isica, Faculdade de Ci\^encias, Unesp-Universidade Estadual Paulista, CP 473, 17033-360 Bauru, SP, Brazil
}
\ead{cjrx@lanl.gov}

\begin{abstract}
  We use atomistic-based simulations to investigate the behavior of ferromagnetic skyrmions being continuously compressed against a rigid wall under dc and ac drives. The compressed skyrmions can be annihilated close to the wall and form a conformal crystal with both a size and a density gradient, making it distinct from conformal crystals observed previously for superconducting vortices and colloidal particles. For both dc and ac driving, the skyrmions can move transverse to the compression direction due to a combination of density and size gradients. Forces in the compression direction are converted by the Magnus force into transverse motion. Under ac driving, the amount of skyrmion annihilation is reduced and we find a skyrmion Magnus ratchet pump. We also observe shear banding in which skyrmions near the wall move up to twice as fast as skyrmions further from the wall. When we vary the magnitude of the applied drive, we find a critical current above which the skyrmions are completely annihilated during a time scale that depends on the magnitude of the drive. By varying the magnetic parameters, we find that the transverse motion is strongly dependent on the skyrmion size. Smaller skyrmions are more rigid, which interferes with the size gradient and destroys the transverse motion. We also confirm the role of the size gradient by comparing our atomistic simulations with a particle-based model, where we find that the transverse motion is only transient. Our results are relevant for applications where skyrmions encounter repulsive magnetic walls, domain walls, or interfaces.
\end{abstract}


\maketitle

\section{Introduction}
Skyrmions are particle like magnetic objects that arise in 
chiral magnets \cite{muhlbauer_skyrmion_2009,yu_real-space_2010,nagaosa_topological_2013}.
Their
size scale, mobility, and stability makes them excellent candidates for
spintronic and logic devices 
\cite{nagaosa_topological_2013,everschor-sitte_perspective_2018,fert_magnetic_2017}.
Due to their gyrotropic properties, skyrmions are also of interest for understanding new types
of collective dynamics in nanostructured systems 
\cite{nagaosa_topological_2013,everschor-sitte_perspective_2018,iwasaki_universal_2013}. 
Many of the proposed applications of skyrmions require them 
to be moved with ac or dc driving,
making it important to identify
possible ways to control the motion and stability of ferromagnetic skyrmions
\cite{pfleiderer_surfaces_2011, wiesendanger_nanoscale_2016,fert_skyrmions_2013,zhang_magnetic_2015,luo_reconfigurable_2018,kang_skyrmion-electronics_2016}. 
Skyrmions have many similarities to other
particle-like objects such as
superconducting vortices, colloids, and
electrons in Wigner crystals. In these systems the 
particle-like objects organize themselves into
triangular lattices, can be set into motion with an external drive,
and can interact with
quenched disorder or pinning \cite{reichhardt_depinning_2016}. 
There are several differences between skyrmions and other overdamped systems. 
Skyrmions experience a strong
non-dissipative Magnus force that can create a velocity component
perpendicular to the external forces, and
the sign of this force
depends
on the topological charge
of the skyrmion \cite{litzius_skyrmion_2017,jiang_direct_2017,lin_driven_2013,lin_particle_2013,iwasaki_universal_2013,nagaosa_topological_2013,zeissler_diameter-independent_2020}.
When subjected to a spin current in a clean sample without defects, 
skyrmions move at an angle, known as the intrinsic 
skyrmion Hall angle $\theta_\text{sk}^\text{int}$ 
\cite{litzius_skyrmion_2017,jiang_direct_2017,lin_driven_2013,lin_particle_2013,iwasaki_universal_2013,nagaosa_topological_2013}, to the applied
driving force.
Experimentally observed skyrmion Hall angles
span the range
from a few degrees up to $50^{\circ}$, depending on the system parameters and the skyrmion size \cite{jiang_direct_2017,zeissler_diameter-independent_2020,litzius_skyrmion_2017,reichhardt_depinning_2016,brearton_deriving_2021},
and 
much larger angles should be possible \cite{nagaosa_topological_2013,reichhardt_depinning_2016}. 
Skyrmions with 
different sizes can be stabilized in a given sample, and
the gyrotropic effects are stronger for smaller skyrmions and
weaker for larger ones,
meaning that skyrmions of different size move with
different
intrinsic skyrmion Hall angles. \cite{zeissler_diameter-independent_2020,litzius_skyrmion_2017,fert_magnetic_2017}.

There is increasing interest in identifying ways to control individual and collective skyrmion motion.
Possible methods include periodic pinning \cite{reichhardt_quantized_2015,reichhardt_nonequilibrium_2018,feilhauer_controlled_2020,vizarim_skyrmion_2020,vizarim_skyrmion_2020a,vizarim_topological_2020,vizarim_shapiro_2020,vizarim_soliton_2022,reichhardt_commensuration_2022},
ratchet effects \cite{reichhardt_magnus-induced_2015,gobel_skyrmion_2021,souza_skyrmion_2021,chen_skyrmion_2019,ma_reversible_2017,chen_ultrafast_2020},
interface guided motion \cite{vizarim_guided_2021,zhang_edge-guided_2022}, 
 strain, magnetic or temperature gradients
 \cite{yanes_skyrmion_2019,zhang_manipulation_2018,everschor_rotating_2012,kong_dynamics_2013}, 
one-dimensional potential wells \cite{purnama_guided_2015}, and skyrmion-vortex coupling using a 
 ferromagnet-superconductor heterostructure \cite{menezes_manipulation_2019}.
Skyrmions can also be manipulated by being
compressed against a wall or linear obstacle, such as by applying
a drive that forces the skyrmions to move toward an interface or extended
nanostructure.
Another possibility would be
to move some form of domain wall against the
edge of the skyrmion lattice.

\begin{figure}
  \centering
  \includegraphics[width=0.6\columnwidth]{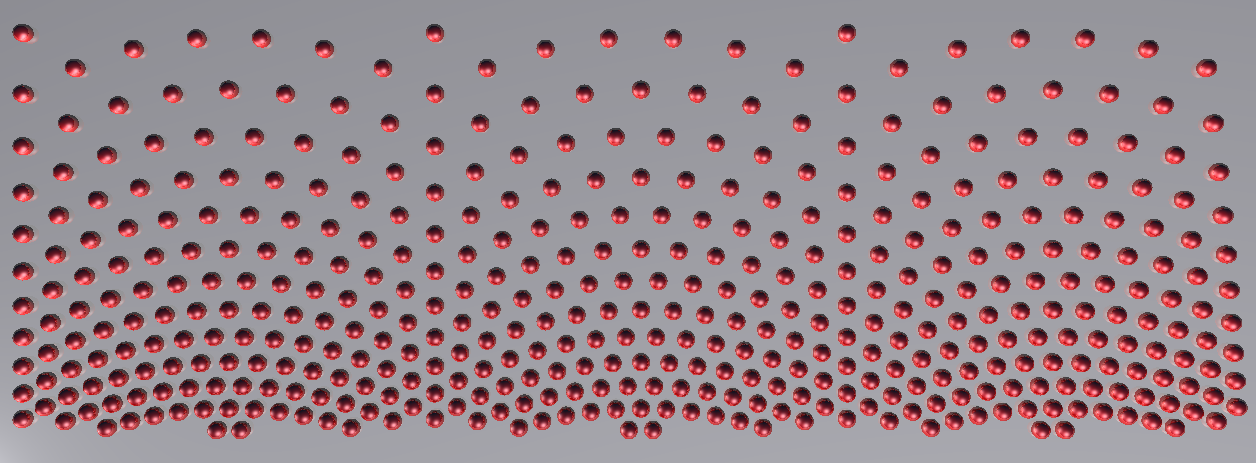}
  \caption{A conformal crystal created by applying a conformal transformation
to a semiannular section of a regular triangular lattice.}
\label{fig1}
\end{figure}

The behavior of
overdamped particles such as
superconducting vortices and colloids under compression
has been examined previously. In a uniform crystal,
the spacing between particles is fixed, so under compression
it can be difficult for the sample to reconfigure into a new uniform
crystal with a smaller lattice constant.
Studies of magnetically
interacting particles being pushed toward a wall showed that the particles
form a density gradient composed of an
array of arcs in a shape known as a conformal crystal
\cite{rothen_conformal_1993,rothen_mechanical_1996}. 
A perfect conformal crystal is produced by conformally transforming
a two-dimensional uniform lattice.
In the transformed structure, a density gradient is present but
the angles between nearest neighbors
in the original lattice are preserved and there are no 
topological defects. The experiments
in Refs.~\cite{rothen_conformal_1993,rothen_mechanical_1996}
produced a conformal
structure containing some topological defects. In figure~\ref{fig1}
we show an example of a perfect two-dimensional conformal crystal 
without defects.
Simulations of superconducting
vortices interacting with pinning sites that have a conformal
crystal arrangement showed that such a substrate produces
optimal pinning in samples where the vortices
enter from one side and develop a natural density gradient
\cite{ray_strongly_2013}, a result that was
then confirmed in experiment \cite{wang_enhancing_2013}.

Menezes {\it et al.} numerically studied
a crystal of type II superconducting vortices,
which have softer intermediate-range interactions than the magnetic
particles, 
being pushed against a wall by a current
\cite{menezes_conformal_2017}.
They showed that under these conditions, a
conformal vortex lattice forms spontaneously.
Further studies explored conformal structures in circular geometries
\cite{menezes_self-assembled_2019}
and with different types of defects \cite{meng_defects_2021}.
These studies focused only on 
the final static conformal states and did not
address the dynamics during compression.

Since magnetic skyrmions
also organize themselves in a triangular array and have 
soft repulsive interactions,
it should be possible to form a conformal crystal by compressing
a skyrmion lattice,
similar to observations for
colloidal particles and superconducting vortices.
In overdamped systems,
the particles move in the direction of the net force they
experience, so application of a dc or ac compression
is not expected to produce
motion transverse to the compression direction.
In the case of skyrmions, however, 
the Magnus force dominates the dynamics and generates a velocity
component perpendicular to the net applied force.
Additionally, 
previous studies involving
conformal crystals were performed with stiff particles
that have no internal degrees of freedom; however,
skyrmions can change size in response to their environment,
and the dynamics of the skyrmion
are affected by the size of the skyrmion.

In previous work, Zhang {\it et al.} \cite{zhang_structural_2022} examined
skyrmions under compression against a wall and focused on structural
transitions in the skyrmion lattice.
No conformal arrangement appeared in this work, however,
since only a portion of the skyrmions were being 
compressed while the remaining skyrmions served as pinning obstacles.
Zhang {\it et al.} 
showed that a lattice
structural transition can be induced by compression, and that
near the injection boundary separating the driving and compressive
regions, the skyrmions may flow in chains. 

In this work, we perform atomistic simulations
of the dynamics of multiple skyrmions being compressed against a rigid wall 
by dc or ac drives
applied perpendicular to the wall. 
This work is distinct from that of Ref.~\cite{zhang_structural_2022}
since we apply the drive to all of the skyrmions,
which should be a
realistic situation for experiments on
skyrmions moving against the edge of a nanofabricated structure.
We show that the skyrmions accumulate near the wall and
assemble into a conformal crystal, and we study the dynamical
formation of this structure.
The skyrmions gradually change in size
from larger to smaller as they approach the wall,
leading to a size gradient 
in the conformal crystal, different from what is observed  
for colloids and superconducting vortices.
Under dc driving, after the skyrmions have formed a compressed conformal
crystal, 
they all move as a jammed packing
in the $-x$ direction.
In addition, the constant
compressing current
leads to constant skyrmion annihilation. 

Under an ac rectified drive,
we still observe
an initial compression of the skyrmion lattice followed
by transverse motion of the skyrmions
along the $-x$ direction.
In addition, 
there is a clear shear banding in which
skyrmions closer to the wall move
much faster than skyrmions far from the wall.
The skyrmion annihilation is reduced due to a relaxation time introduced
by the ac driving.
When the current is in the off state,
the compression of the skyrmion lattice is relieved, enabling a larger
number of skyrmions to be sustained inside the sample over longer
periods of time compared to what we observe under dc driving.

We also confirm that the
skyrmion size gradient and the resulting spatial gradient
in the skyrmion Hall angle is crucial for
producing the transverse motion. We compare our atomistic results 
with simulations of
a particle-based model of rigid skyrmions that have fixed size and
fixed intrinsic skyrmion Hall angle. In the particle-based model,
the motion in the transverse direction is transient 
and vanishes as time evolves.

\section{Simulation}
We simulate a ferromagnetic thin film that supports the presence of Neél skyrmions
at $T = 0$ K
under a magnetic field of magnitude
$0.2 D^2 / J < H < 0.8  D^2 / J$ applied perpendicular to 
the film plane, 
$\mathbf{H}=(0,0,H)$.
Here $D$ is the strength of the Dzyaloshinskii-Moriya interaction and
$J$ is the strength of the exchange interaction.
The film has infinite dimensions in
the $x$ and $y$ directions, but we introduce a rigid wall at $y=0$.  
We use atomistic model simulations \cite{evans_atomistic_2018} to stabilize a skyrmion lattice in
the sample
with the following Hamiltonian \cite{iwasaki_current-induced_2013, iwasaki_universal_2013, seki_skyrmions_2016}:

\begin{eqnarray}\label{Eq1}
\fl\mathscr{H}=-\sum_{i, j\in N}J_{ij}\mathbf{m}_i\cdot\mathbf{m}_j
               -\sum_{i, j\in N}\mathbf{D}_{ij}\cdot\left(\mathbf{m}_i\times\mathbf{m}_j\right)
              -\sum_{i}\mu\mathbf{H}\cdot\mathbf{m}_i \nonumber\\
              -\sum_{i\notin W}K_1\left(\mathbf{m}_i\cdot\hat{\mathbf{z}}\right)^2
              -\sum_{i\in W}K_2\left(\mathbf{m}_i\cdot\hat{\mathbf{z}}\right)^2 
\end{eqnarray}

The first term on the right hand side is the exchange interaction between adjacent spins $\mathbf{m}_i$ and 
$\mathbf{m}_j$,
the second term is the interfacial Dzyaloshinskii-Moriya interaction in the thin film,
the third term is the Zeeman interaction due to the application of the magnetic field,
where $\mu$ is the magnitude of the local atomic spin moment,
and the last two terms are the out of plane magnetic anisotropies
\cite{iwasaki_universal_2013, iwasaki_current-induced_2013, zhang_antiferromagnetic_2016, stosic_pinning_2017} consisting of
the sample anisotropy (fourth term) and
the wall anisotropy (fifth term).
%
We only consider interactions with the first neighbors, given by the set
$N$.
The wall at which the skyrmion lattice is compressed is modeled using magnetic uniaxial anisotropy 
defects at $y=0$, where we set
$K_2=5J$.
The spins composing the wall are in the set $W$.

We apply a spin transport current
along the $-y$ direction in order to push
the skyrmions towards the wall.
The spin time evolution is given by the LLG equation \cite{iwasaki_current-induced_2013, iwasaki_universal_2013, seki_skyrmions_2016}:

\begin{equation}\label{Eq2}
        \frac{d\mathbf{m}_i}{dt}=
        -\gamma\mathbf{m}_i\times\mathbf{H}^\text{eff}_i
        +\alpha \mathbf{m}_i\times\frac{d\mathbf{m}_i}{dt}
        +\frac{pa^3}{2e}\left(\mathbf{j}\cdot{\nabla}\right)\mathbf{m}_i
\end{equation}

Here $\gamma$ is the gyromagnetic ratio given by $\gamma=g\mu_B/\hbar$, $\alpha$ 
is the Gilbert damping parameter, $\mathbf{H}^\text{eff}_i=-\frac{1}{\mu_i}\frac{\partial \mathscr{H}}{\partial \mathbf{m}_i}$ is the effective field, $\mu_i$ is the local spin magnetic moment, and 
the last term corresponds to the applied current, where $p$ is the polarization, $e$ is the electron 
charge, and $a$ is the lattice constant.
This is called the spin-transfer-torque term, and it includes the assumption
that the conduction electron spins are parallel to the local magnetic moments $\mathbf{m}$ \cite{iwasaki_universal_2013, zang_dynamics_2011}.

We consider two different types of applied current.
The first is a dc applied current with
$\mathbf{j}=-j_0 \mathbf{\hat{y}}$.
The second is rectified ac driving given by:

\begin{equation}\label{Eq3}
    \mathbf{j}=\left\{\begin{array}{cc}
        -j_0\sin\left(2\pi\omega t\right)\hat{\mathbf{y}} & \text{for}\sin\left(2\pi\omega t\right) > 0\\
        0 & \text{for} \sin\left(2\pi\omega t\right) \leq 0
    \end{array}\right.
\end{equation}

where $t$ is time and $\omega$ is the oscillation frequency.
This ac current consists of two stages,
one with $\mathbf{j} \neq 0$,
referred to throughout the work as the \textit{compression cycle},
and one in which the drive is absent,
referred to as the \textit{relaxation cycle}.

Since skyrmions carry topological charge, in order to quantify the number of skyrmions in the sample, we 
measure the net topological charge in the sample $Q$, given in the continuum
limit by
\cite{iwasaki_universal_2013, zhang_antiferromagnetic_2016, gobel_skyrmion_2021, zhang_magnetic_2015, kim_quantifying_2020, seki_skyrmions_2016}

\begin{equation}
    Q_\text{cont} = \frac{1}{4\pi}\int\mathbf{m}\cdot\left(\frac{\partial \mathbf{m}}{\partial x}\times\frac{\partial \mathbf{m}}{{\partial y}}\right)dx dy \ .   
\end{equation}

We are performing atomistic simulations, and therefore the discrete nature of the atomic lattice must be taken into account by computing the skyrmion charge with a discrete sum instead of an integral:
\begin{equation}
    Q_\text{disc} = \frac{1}{16\pi}\sum_{i, j}\mathbf{m}_{i, j}\cdot\left[\left(\mathbf{m}_{i, j+1}-\mathbf{m}_{i, j-1}\right)\times\left(\mathbf{m}_{i+1, j}-\mathbf{m}_{i-1, j}\right)\right]
\end{equation}
The consequence of this discreteness is that the charge of an individual
skyrmion is not always
equal to $\pm1$.
Instead, it depends on the skyrmion size,
and larger skyrmions have $Q$ closer to $\pm1$ \cite{kim_quantifying_2020}.

We also calculate the skyrmion velocities using the emergent electromagnetic fields \cite{seki_skyrmions_2016}:
\begin{equation}
    E^\text{em}_i=\frac{\hbar}{e}\mathbf{m}\cdot(\partial_i\mathbf{m}\times\partial_t\mathbf{m})
    \;\;,\;\;
    B^\text{em}_i=\frac{\hbar}{2e}\varepsilon_{ijk}\mathbf{m}\cdot(\partial_j\mathbf{m}\times\partial_k\mathbf{m})
\end{equation}
where $\varepsilon_{ijk}$ is the totally anti-symmetric tensor.
The drift velocity $\mathbf{v}_d$ is then computed according to $\mathbf{E}^\text{em}=-\mathbf{v}_d\times\mathbf{B}^\text{em}$ \cite{seki_skyrmions_2016, schulz_emergent_2012}.
From the velocities we determine the skyrmion displacement as 
$\Delta \mathbf{r}=\int\mathbf{v}_ddt$.

Throughout the simulation we fix $\alpha=0.3$, $p=-1.0$ and $a=5$\AA.
Unless otherwise noted, we
assume the following values for the material parameters:
$J=1\text{meV}$, $D=0.5J$, $\mathbf{H}=0.5\left(D^2/J\right)\hat{\mathbf{z}}$, $K_1=0.05J$, and
 $\omega=3\times10^9\text{Hz}$.

We use the Runge-Kutta fourth order integration method. 
The integration is done by normalizing the time in dimensionless units $t=(\hbar/J)\tau$. 
The current is also normalized in dimensionless units
$\mathbf{j}=(2eJ/a^2\hbar)\mathbf{j}'$.

\section{Compression and Dynamics Under a dc drive}

\begin{figure}[h]
    \centering
    \includegraphics[width=0.3\columnwidth]{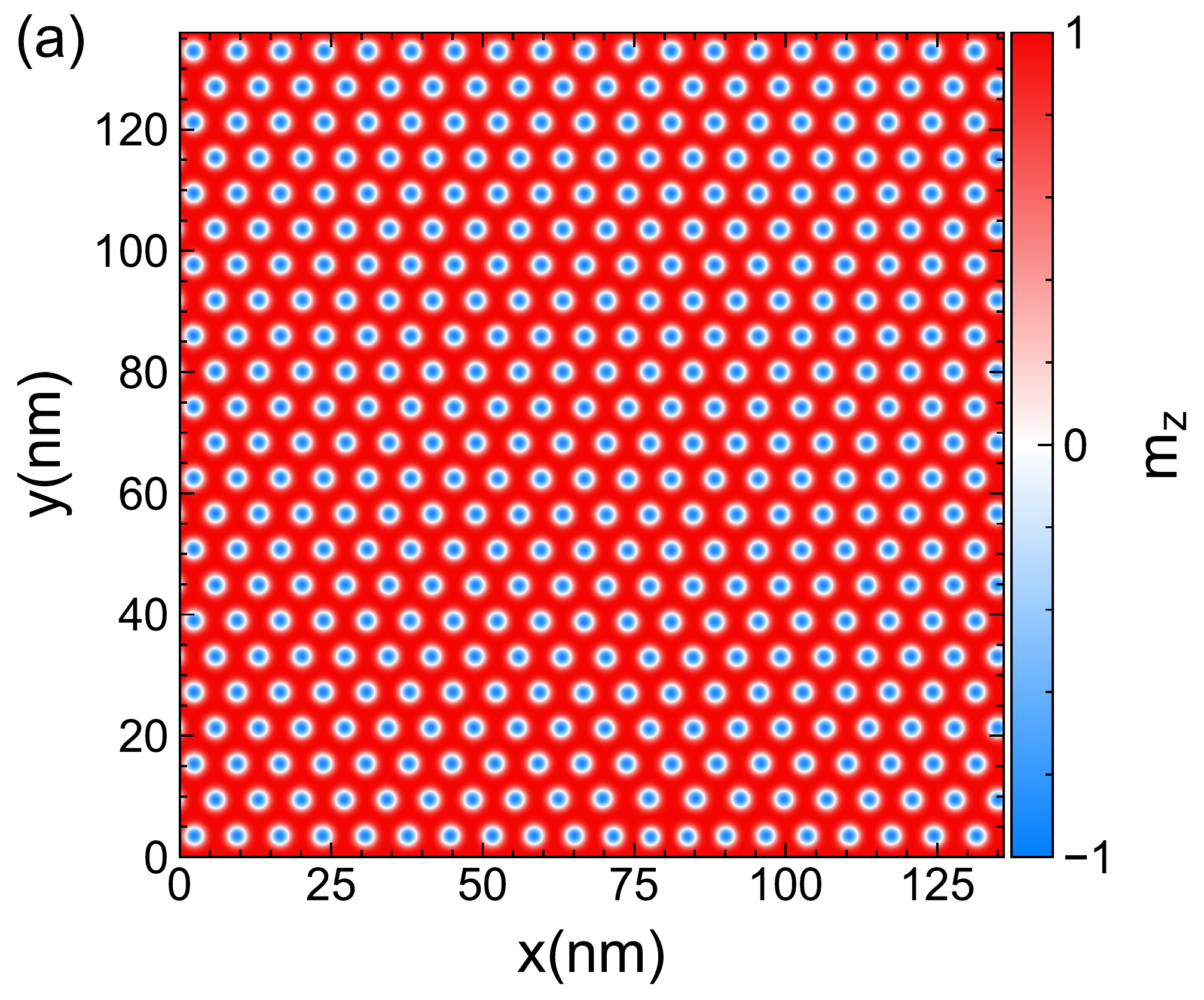}
    \includegraphics[width=0.3\columnwidth]{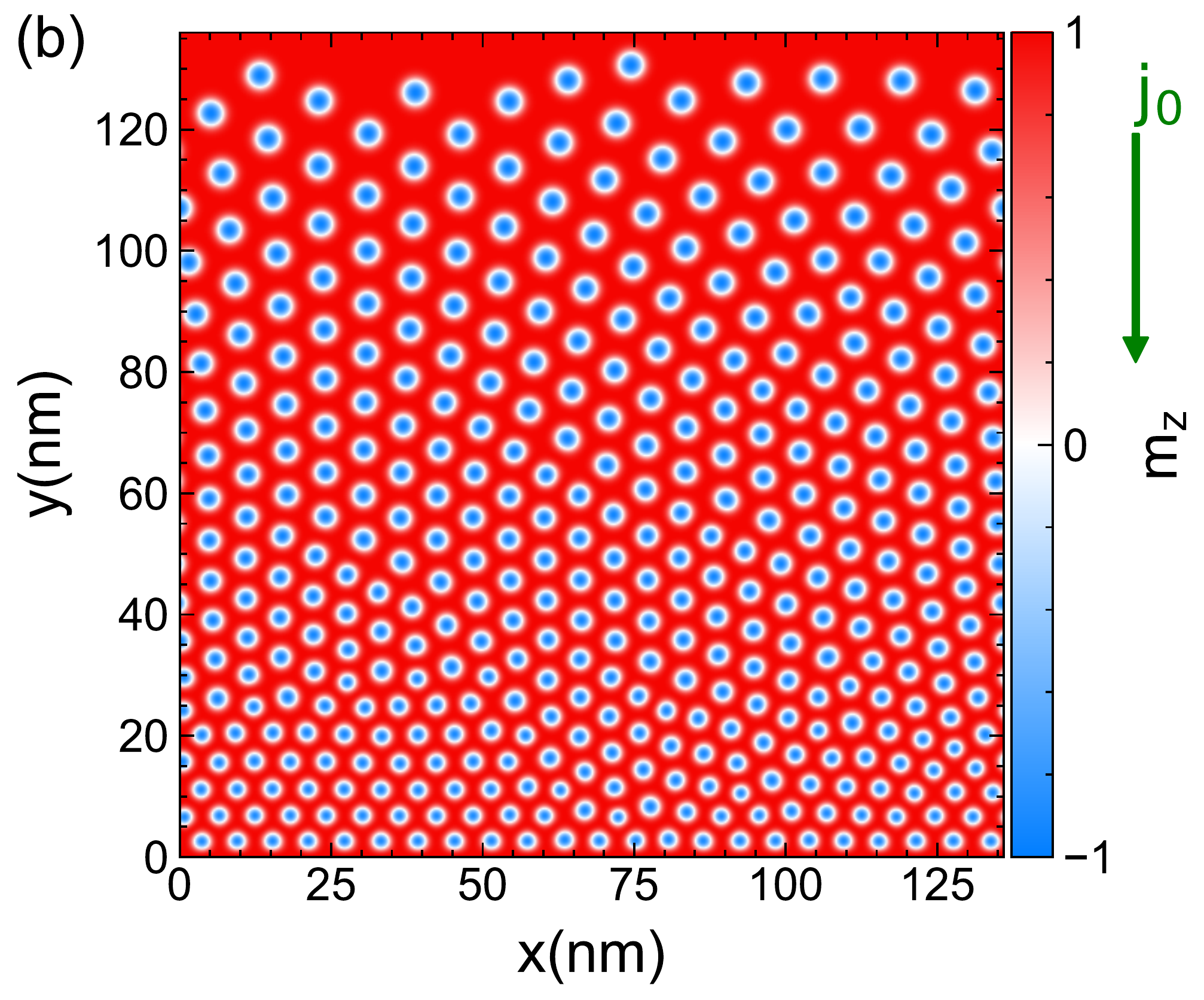}
    \includegraphics[width=0.3\columnwidth]{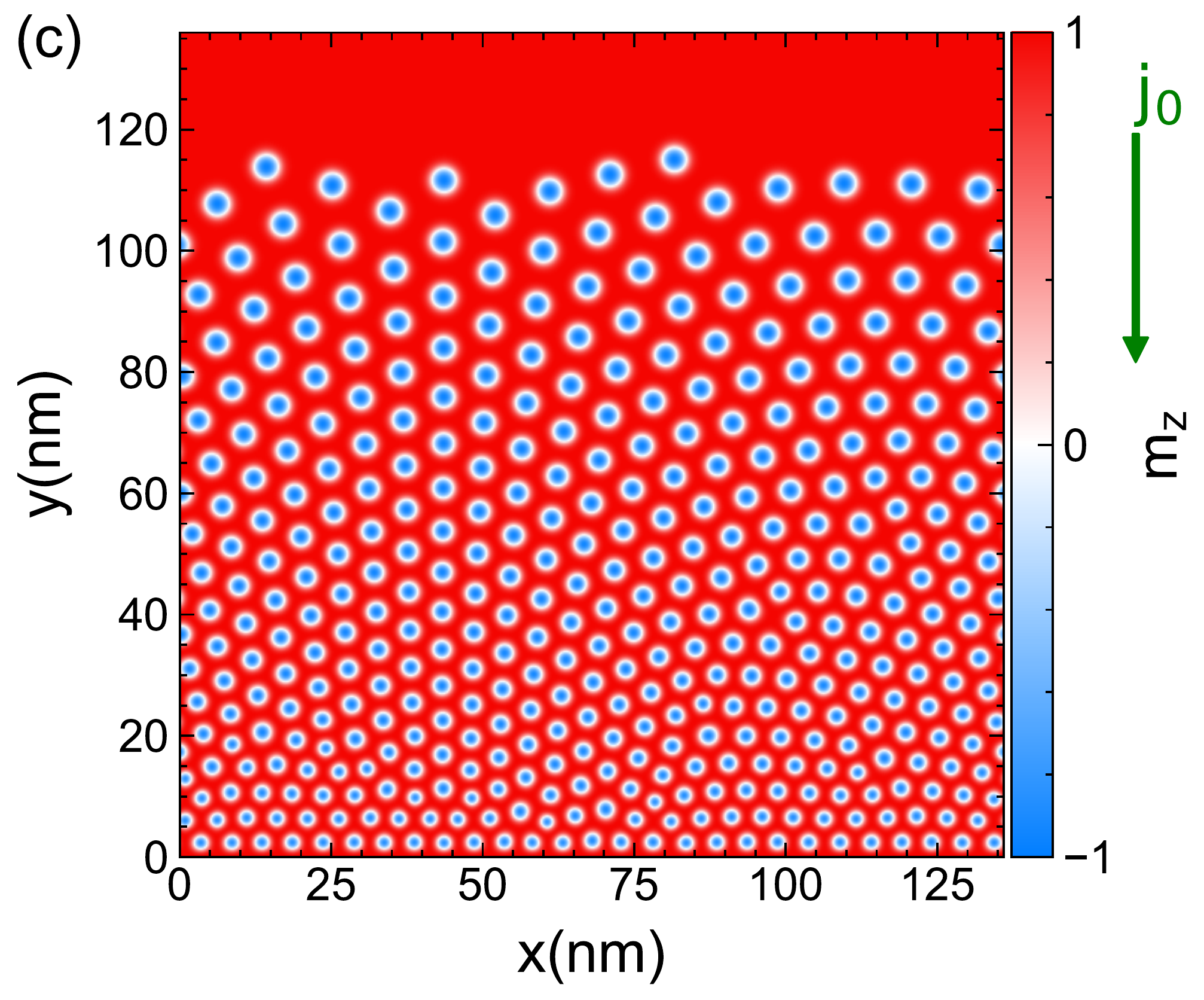}
    \caption{
      Images of the spin configurations at different times under dc driving
      with $j_0 = 3.2\times10^{9}\text{A/m}^2$.
    (a) The skyrmion 
      lattice for $t = 0$ ns in the absence of an external drive,
      where the skyrmions organize themselves into a 
      triangular lattice.
      (b) 
      At $t = 39.5$ ns,
      the skyrmions form
      a moving amorphous crystal.
    (c) At $t = 197.5$ ns,
      the skyrmions
      form a moving conformal 
      crystal containing a size gradient with smaller skyrmions
      located closer to the wall.
    }
    \label{fig2}
\end{figure}

We
first consider the
case of skyrmions being pushed towards the wall with a fixed dc
spin current $\mathbf{j}=-j_0\mathbf{\hat{y}}$, where $j_0 = 3.2\times10^{9}\text{A/m}^2$.
The system is initialized by
stabilizing a skyrmion lattice in the absence of a
spin current using an applied magnetic field 
$\mathbf{H}=0.5\left(D^2/J\right)\hat{\mathbf{z}}$.
As can be seen in figure~\ref{fig2}(a),
skyrmions organize themselves in a perfect triangular lattice as expected. As the external spin
current is applied along the $-y$ direction, the skyrmions begin to move and the lattice starts to
compress against the wall, as shown in figure~\ref{fig2}(b).
The perfect triangular orientation of the lattice is destroyed
due to the stress induced by the external drive.
As the drive becomes stronger, the skyrmion lattice stabilizes and
reorganizes into a conformal lattice
structure as shown in figure~\ref{fig2}(c).
This structure
is very similar to the conformal
vortex crystals found by Menezes and de Souza Silva in type II
superconductors \cite{menezes_conformal_2017}.
In a perfect conformal crystal, illustrated
in figure~\ref{fig1}, there are no defects;
however, in both the magnetic colloid and
superconducting vortex systems 
\cite{rothen_conformal_1993,menezes_conformal_2017},
defects are present, similar to what we find
in figure~\ref{fig2}(c).
The wavelength of the arch structure in the conformal crystal
decreases as the amount of compression increases, so in a
sample of the size we consider, multiple arches appear if we
increase the compression beyond what is shown in figure~\ref{fig2}(c).
Two aspects of the skyrmion conformal crystal structure
differ from previous conformal crystal observations.
First,
the crystal is not static but is moving in the $-x$
direction, transverse to the compression.
Second, we find
a skyrmion size gradient in which
skyrmions near the wall are smaller while 
skyrmions far from the wall are larger.

\begin{figure}[h]
    \centering
    \includegraphics[width=0.3\columnwidth]{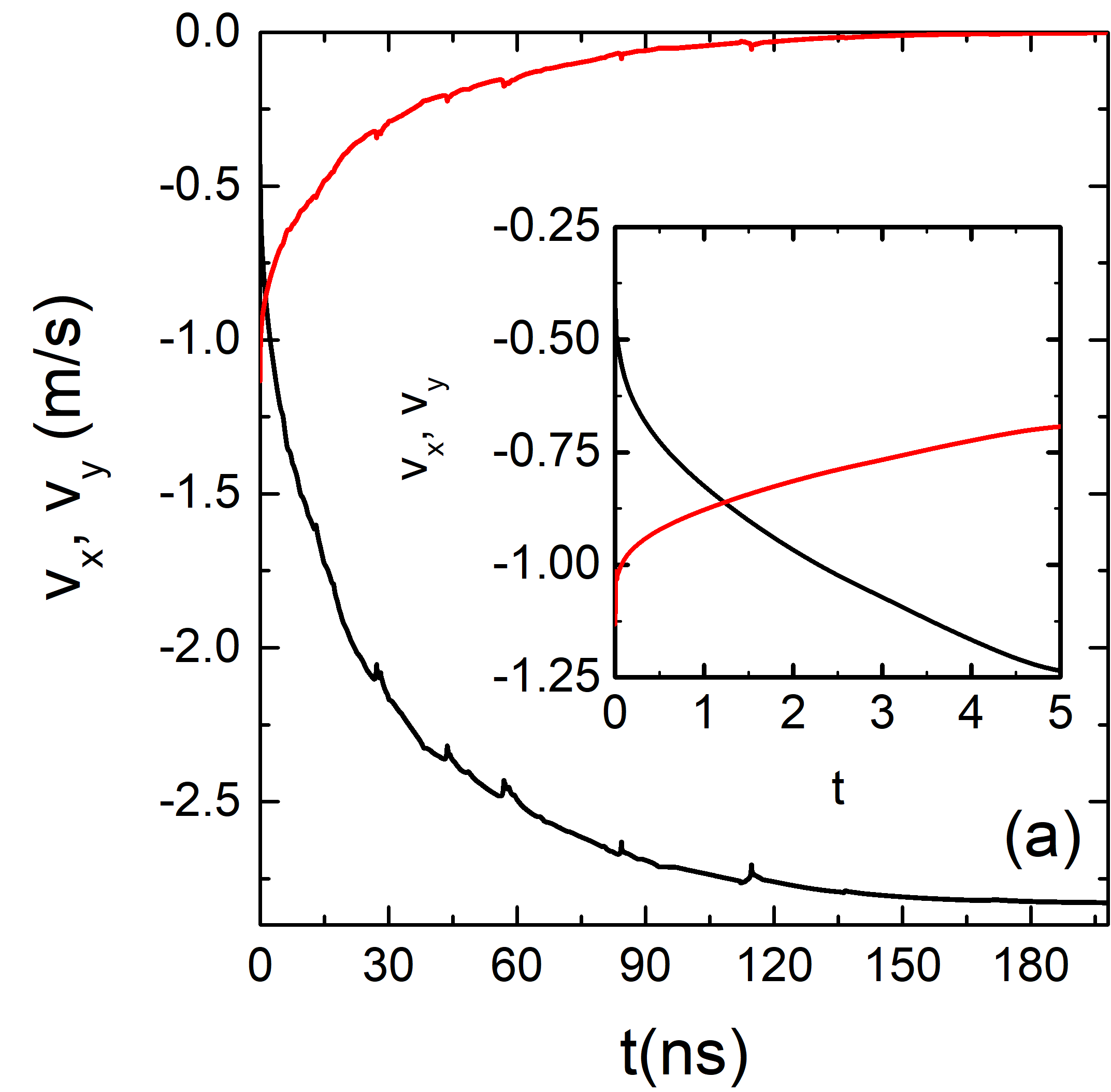}
    \includegraphics[width=0.3\columnwidth]{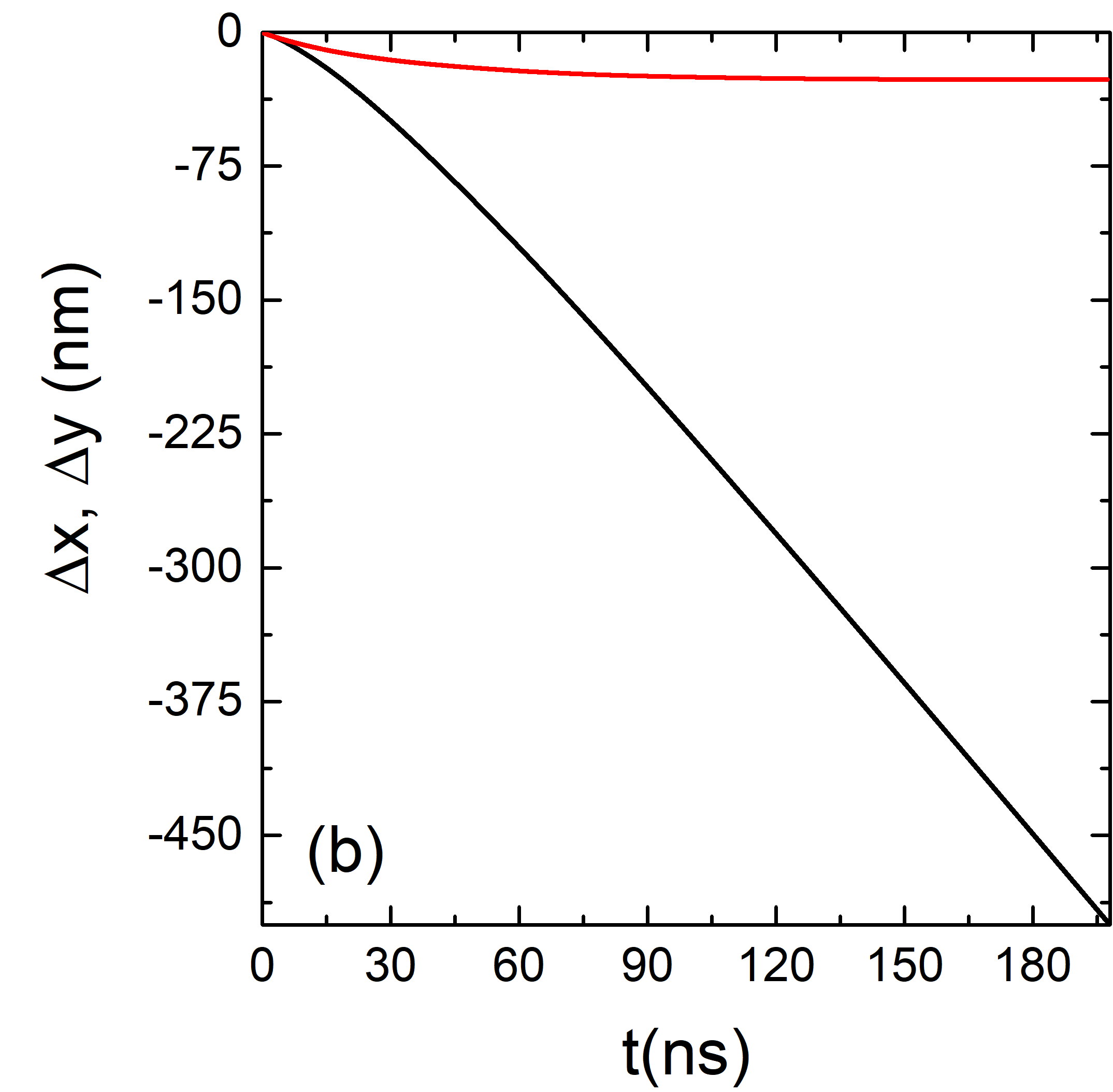}
    \includegraphics[width=0.3\columnwidth]{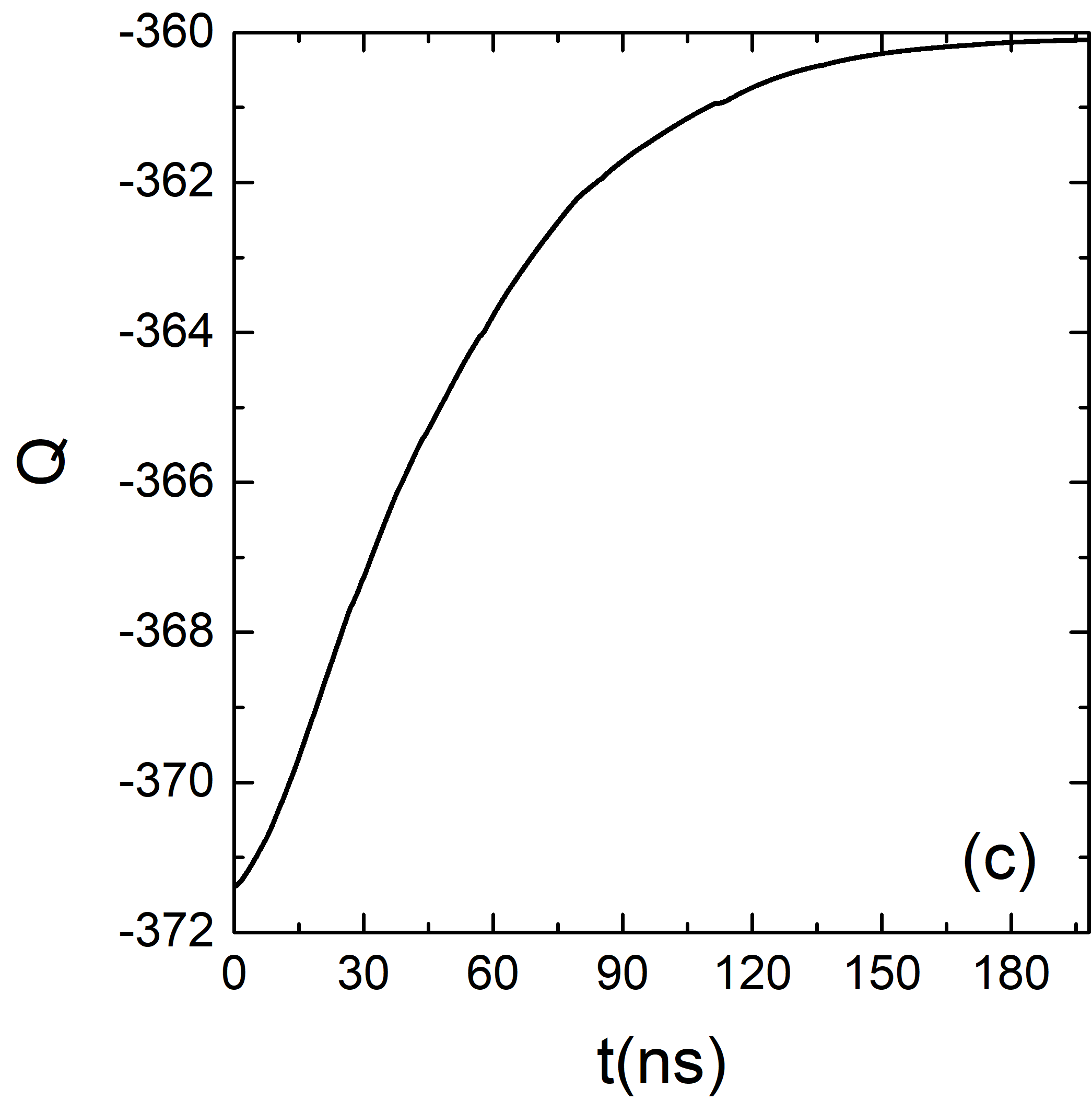}
    \caption{Response as a function of time
      in a sample with dc driving of $j_0=3.2\times10^{9}\text{A/m}^2$. 
      (a) Skyrmion velocity components $v_x$ (black) and $v_y$(red).
      Inset: zoomed in view of the early time behavior from the main panel.
    (b) Net motion of the skyrmion lattice
    perpendicular ($\Delta x$, black) and parallel ($\Delta y$, black) to
    the compression.
    The lattice reaches its maximum compression after 
    $90 \text{ns}$, the point at which $\Delta y$ attains its saturation value.
    (c) The topological charge $Q$ vs time.}
    \label{fig3}
\end{figure}

To illustrate the motion of the skyrmion conformal crystal,
in figure~\ref{fig3}(a) we show the average
skyrmion velocity components $v_x$ and $v_y$ as a function of time.
We obtain $v_x(t)$ and $v_y(t)$ by summing the
instantaneous velocity of all skyrmions
and dividing by the number of skyrmions
present at time $t$.
The external drive $\mathbf{j}$
is applied beginning at $t=0$ ns.
As shown in the inset of figure~\ref{fig3}(a), $|v_y|>|v_x|$
for $t<1.2$ ns. 
During this time period,
the external drive pushing skyrmions in the $-y$ direction
dominates the dynamics.
Skyrmion-wall and skyrmion-skyrmion interactions become important for
$t>1.2$ ns
and the
dynamics changes drastically,
with $v_x$ increasing in magnitude
and $v_y$ decreasing in magnitude exponentially until reaching 
saturation values of $v_x = -2.8$ m/s and $v_y = 0$ m/s,
respectively.
At saturation,
the skyrmions have
organized into the conformal lattice illustrated in
figure~\ref{fig2}(c) and move with a constant velocity in the
$-x$ direction.
This is different than what was observed in
the superconducting vortex systems
considered by Menezes and de Souza Silva
\cite{menezes_conformal_2017},
where the conformal vortex lattice is static.
There are some small spikes
in $v_x$ and $v_y$ in figure~\ref{fig3}(a)
that are correlated 
with skyrmion annihilation
events or structural rearrangements
that occur as the skyrmions are compressed. 
In figure~\ref{fig3}(b) we plot the skyrmion
displacements $\Delta x$ and $\Delta y$
with respect to the $t=0$ positions, and find
that $\Delta x$ continuously increases with time,
while $\Delta y$
saturates at a maximum displacement of $\Delta y^{max} = -26.5$ nm.
This is the maximum
compression that the lattice can endure in the $y$ direction due to the combination of external drive,
skyrmion-skyrmion, and skyrmion-wall interactions. 
Figure~\ref{fig3}(c) shows the topological
charge $Q$ in the sample as a function of time. 
This measurement allows us to quantify the change in the number and size
of skyrmions in the sample.
Higher values of $Q$ indicate that more and/or larger skyrmions
are present, while lower values indicate
that the number and/or size of the skyrmions has been reduced.
We find a nonmonotonic behavior in the rate of change of $Q$,
with an increasingly rapid reduction in $|Q|$ with
time
for $t<30$ ns, 
a steady reduction in $|Q|$ for $30 < t < 90$, and a 
decreasing rate of reduction in $|Q|$
for $t>90$ ns.
At long times, 
the topological charge stabilizes around $Q= -360$,
indicating that the size and number of skyrmions reaches a steady value.

A simple explanation for the
appearance of transverse motion
is that, due to the presence of the Magnus force,
the skyrmion structure can be viewed as a system with odd viscosity,
which creates
flows perpendicular to any pressure gradients \cite{banerjee_odd_2017}.
Additionally there have been several studies investigating 
the enhancement of motion for a skyrmion approaching
a wall,
known as a speed-up effect 
\cite{reichhardt_depinning_2016,iwasaki_colossal_2014,reichhardt_magnus-induced_2015,xing_enhanced_2020,chen_ultrafast_2020,souza_clogging_2022}. 
There have also been experiments on driven skyrmions in which it
was observed that, due to the
skyrmion Hall angle, skyrmions accumulate along the wall
creating a skyrmion density gradient 
\cite{sugimoto_nonlocal_2020}. 
This experimental study was performed in the low density limit,
but we expect that a conformal structure would have
formed if a crystalline lattice had been driven. 
In this work we focus on driving only in
the $-y$ direction. It would also be possible to apply
a current
along the $x$-direction so that
the skyrmion Hall effect drives the skyrmions toward the
wall, also creating a conformal structure.
In this case, motion along the $x$ direction would be
produced directly by the spin current, but there would also
be a speed-up effect from the Magnus force for skyrmions
near the wall.

\begin{figure}[h]
    \centering
    \includegraphics[width=\columnwidth]{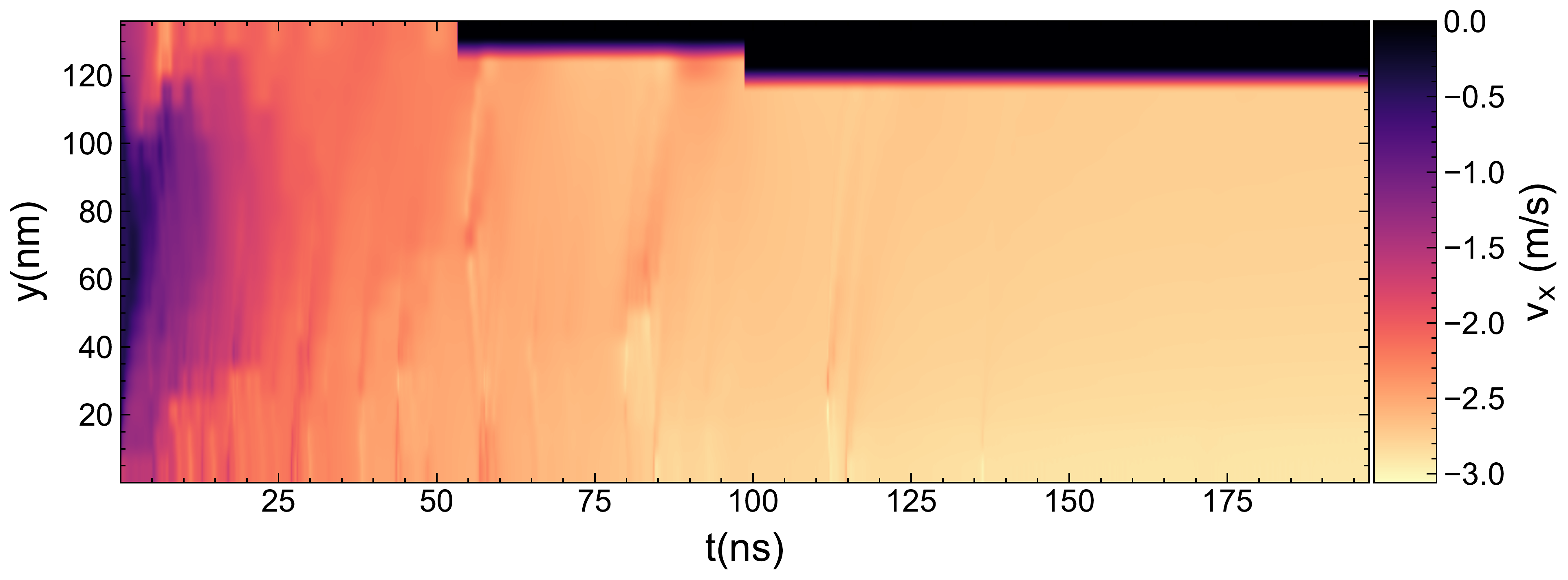}
    \caption{Heatmap of skyrmion velocity $v_x$
    as a function of the distance from the wall $y$ and the time $t$ 
    for dc driving with $j_0=3.2\times10^9\text{A/m}^2$.
    Lighter colors indicate more rapidly moving skyrmions.
    Initially, the skyrmions in the center of the sample are moving
    at the slowest speeds, but once the system reaches a stably compressed
    state, the entire skyrmion assembly moves as a rigid unit.
    The stable compression causes the skyrmions on the bottom
    of the sample to be pushed towards the wall and the magnetic
    pressure causes them to deform and shrink in size.}
    \label{fig4}
\end{figure}

For a better representation of the velocity distribution inside the sample, 
in figure~\ref{fig4} we plot a heatmap of the skyrmion velocity $v_x$ as
a function of the distance from the wall, $y$, and time, $t$.
At early times,
the external drive 
compresses the lattice and tends to move the upper and lower skyrmions 
in a chain of motion.
This behavior is similar to that observed by Zhang {\it et al.}
\cite{zhang_structural_2022}, where a compressing
drive produced chain-like motion
of skyrmions along the boundary between the driven
and non-driven skyrmions.
Here, our simulation shows that the skyrmions in the
central part of the sample are the last to begin moving, which happens
only when the system is completely jammed.
In addition, figure~\ref{fig4}
clearly shows that for
$t>137$ ns all skyrmions move with the same $v_x$, 
indicating that the conformal skyrmion lattice translates
as a rigid object.

\section{Compression under rectified ac drive}

We next consider the effects of applying a rectified ac drive,
where the drive cycle has two stages:
(i) a compressing drive acting on the skyrmions,
which we refer to as the
\textit{compression cycle},
and (ii) the drive is absent and the skyrmion-skyrmion and
skyrmion-wall interactions are dominant,
which we refer to as the \textit{relaxation cycle}.
The spin current is given by equation \ref{Eq3} using $j_0=8.01\times10^9\text{A/m}^2$.

\begin{figure}[h]
   \centering
    \includegraphics[width=0.3\columnwidth]{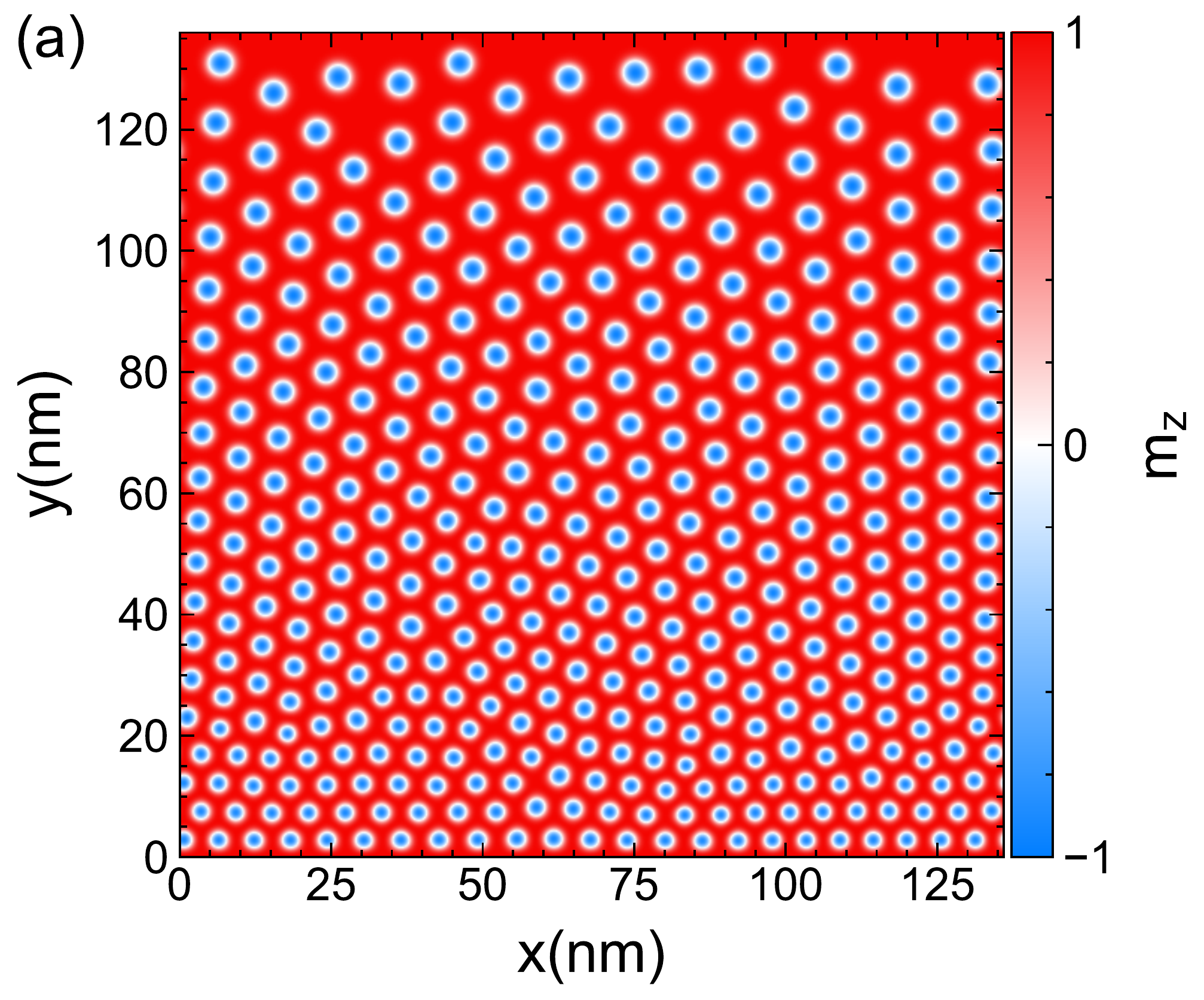}
    \includegraphics[width=0.3\columnwidth]{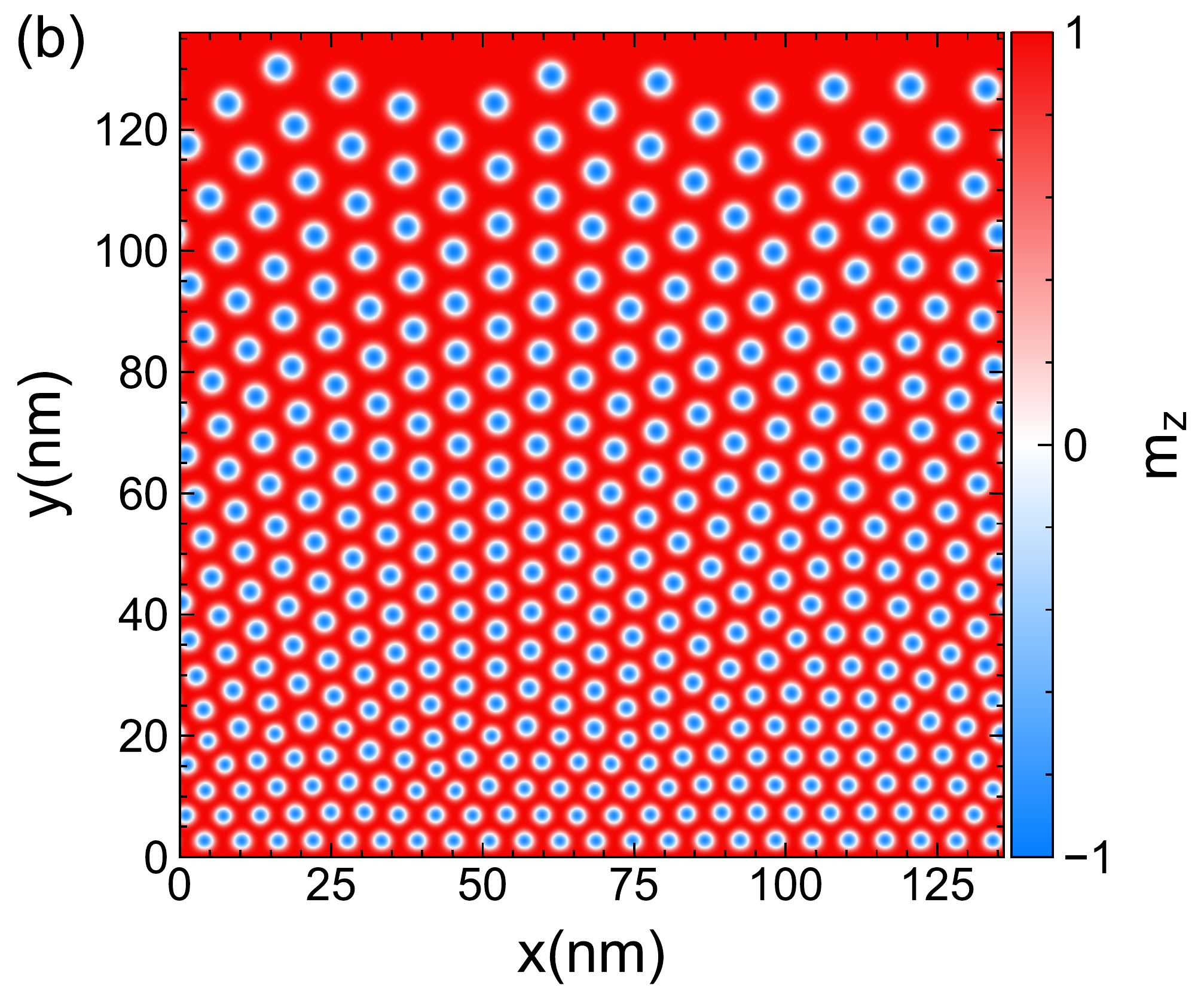}
    \caption{
    Images of the spin configurations at different times under a rectified
    ac drive with  
    $j_0 = 8.01\times10^{9}\text{A/m}^2$. 
    (a)
    At $t = 60$ ns, the skyrmions form a moving amorphous crystal.
    (b) At $t = 100$ ns,
    the skyrmion lattice is compressed into a conformal arrangement,
    but does not become as compressed as what is found for dc driving.
    }
    \label{fig5}
\end{figure}

Figure~\ref{fig5} illustrates some spin configurations
for a system with an applied magnetic
field of $H = 0.5(D^2/J)\mathbf{\hat{z}}$ under rectified ac driving.
The initial configuration at 
$t=0$ ns, in the absence of an applied drive,
is the same as what is shown in figure~\ref{fig2}(a), where the skyrmions
organize into a triangular lattice.
When the ac drive is applied,
the skyrmions begin to be pushed towards the
wall, but due to the alternation of compression and relaxation
cycles, the skyrmions do not become as compressed as in the case of dc
driving.
In figure~\ref{fig5}(a) at $t= 60$ ns, the 
skyrmion lattice is no longer triangular
but has a clear density gradient along the $-y$ direction.
For $t=100$ ns in figure~\ref{fig5}(b),
the skyrmions organize in a compressed configuration
that is partially conformal. Note that this 
is different than what we find for dc driving,
where due to the constant compressing force,
the conformal lattice is more stable.
Under ac driving, the skyrmions are less
tightly packed than in the case of dc driving.
For higher amplitude ac driving,
the skyrmions develop a more strongly conformal structure during
the compression cycle. 

\begin{figure}[h]
    \centering
    \includegraphics[width=0.3\columnwidth]{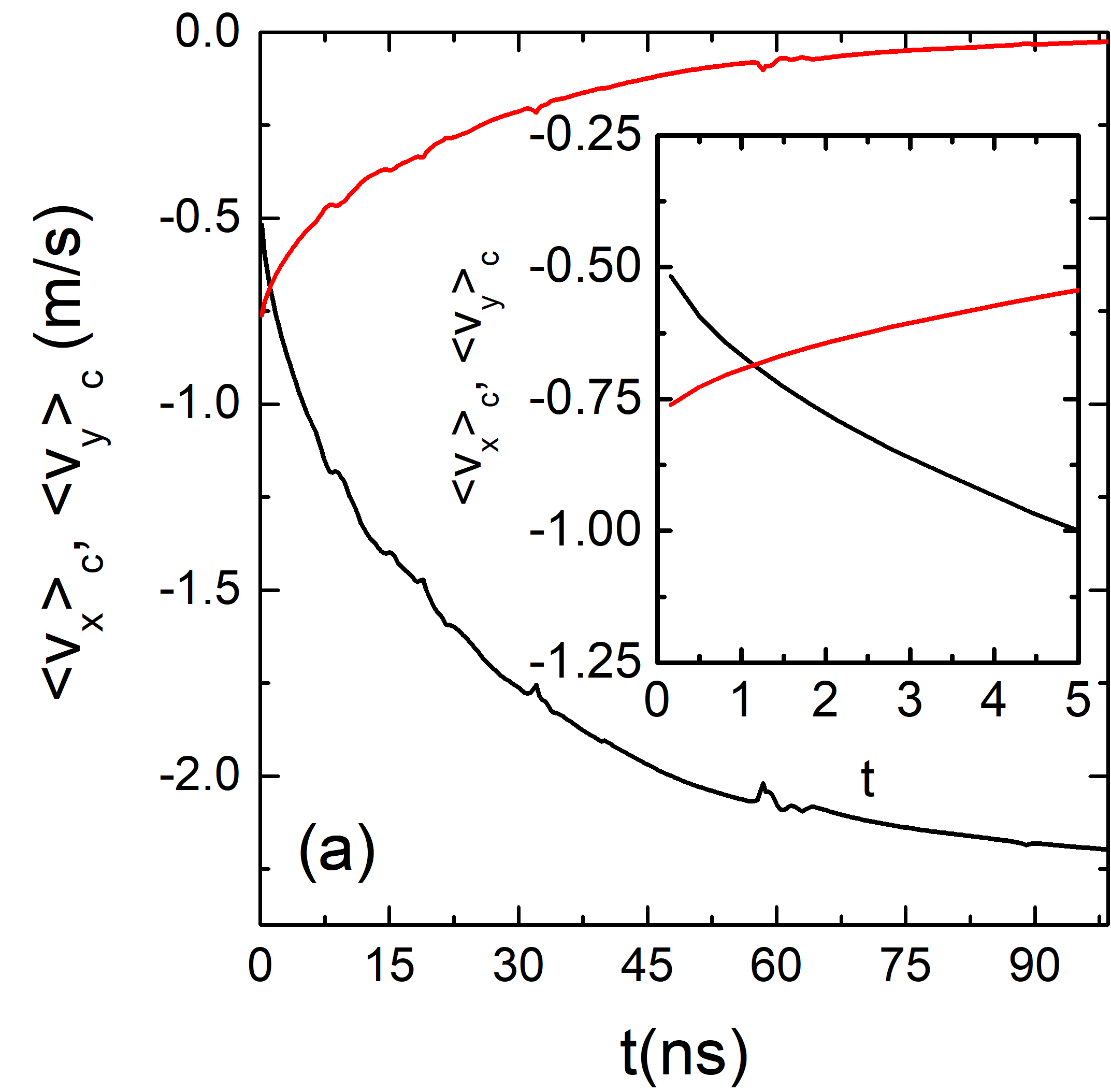}
    \includegraphics[width=0.3\columnwidth]{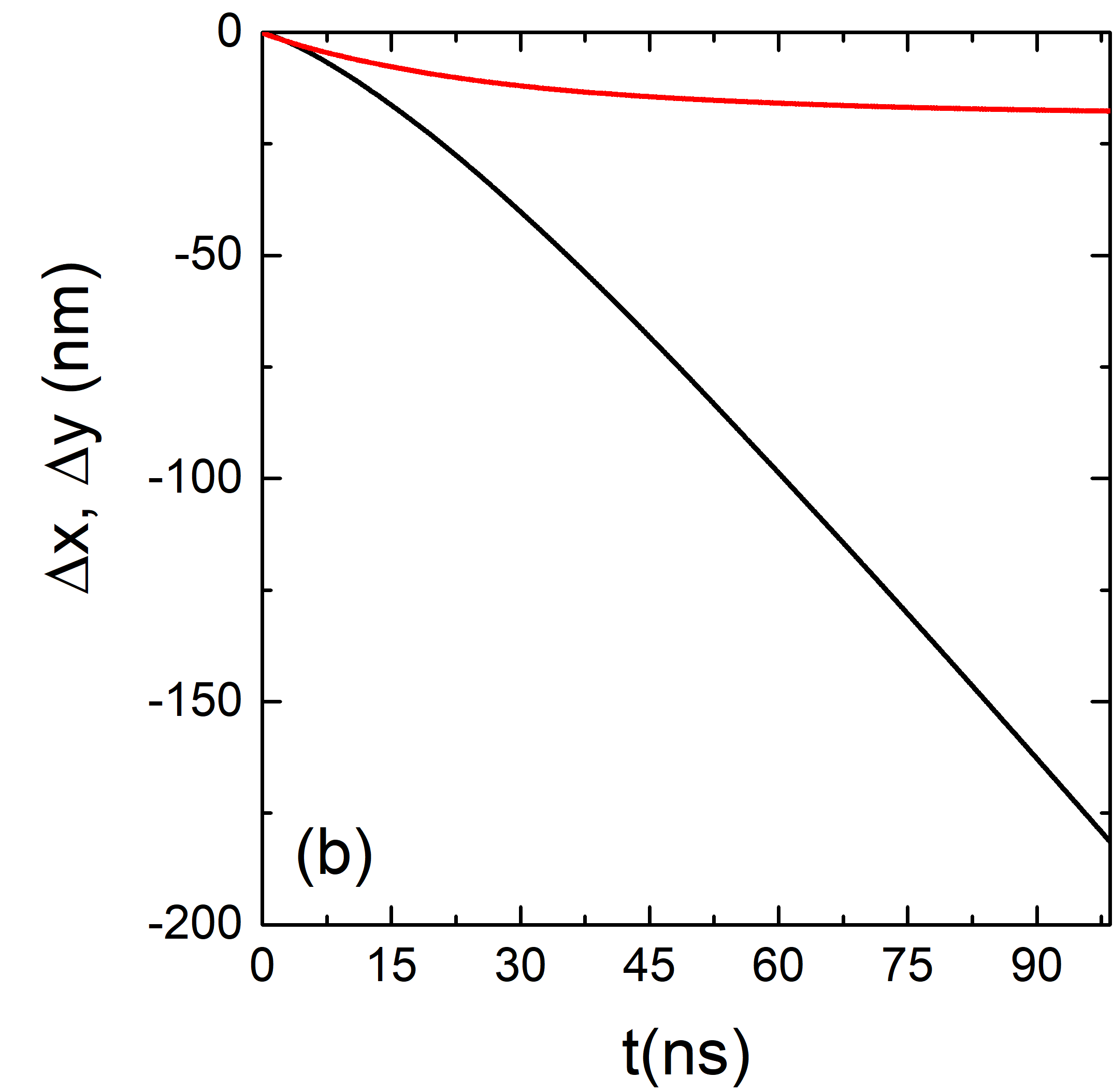}
    \includegraphics[width=0.3\columnwidth]{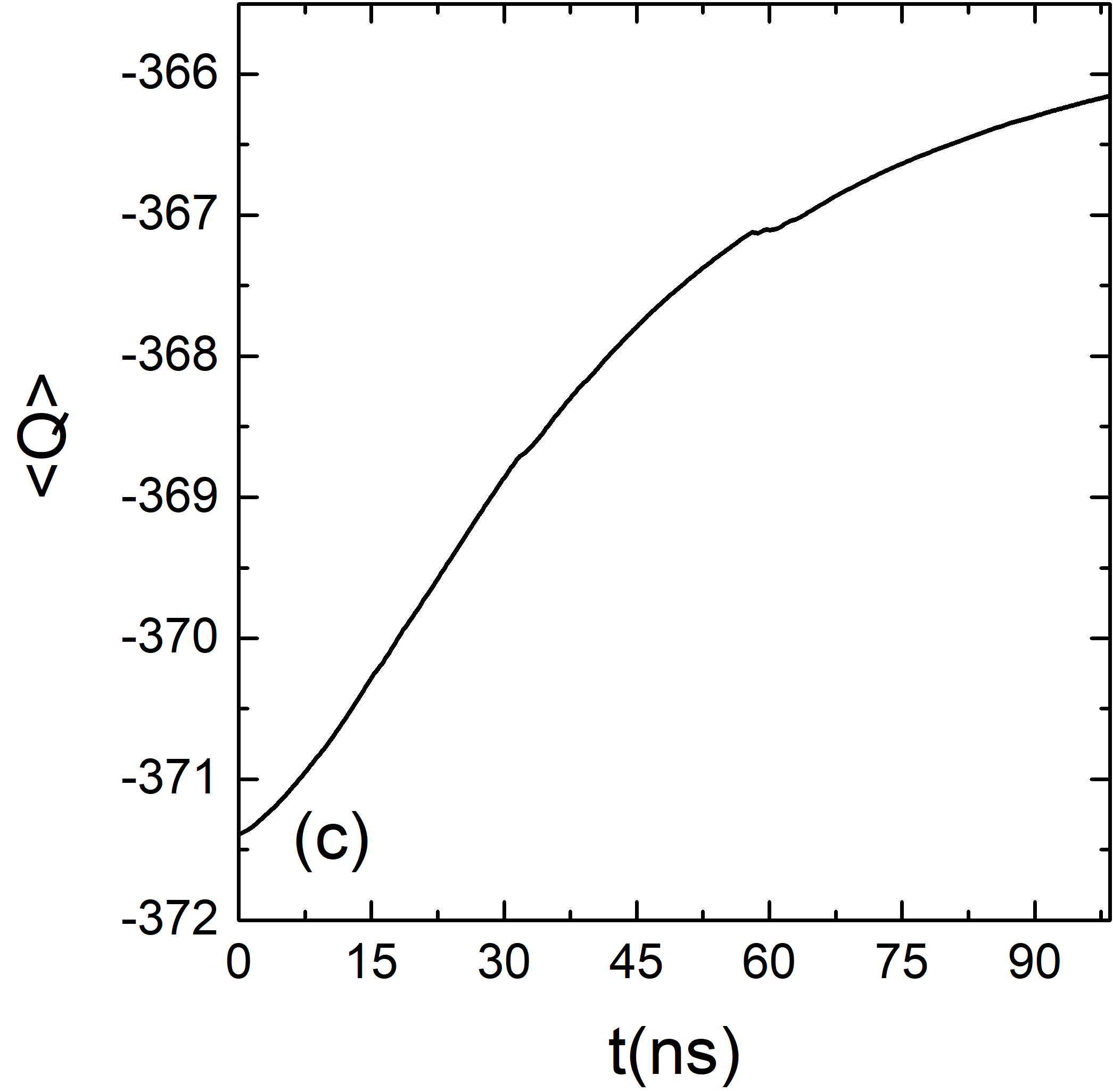}
    \caption{Response as a function of time in a sample under rectified
      ac driving with
$j_0=8.01\times10^{9}\text{A/m}^2$.
      (a) Skyrmion velocity components
      averaged over the ac drive cycle, $\langle v_x \rangle_c$ (black) and
    $\langle v_y \rangle_c$ (red).
      (b) Net motion of the skyrmion lattice perpendicular
      ($\Delta x$, black) and parallel ($\Delta y$, red) to the compression.
      The system reaches its maximum compression displacement
      at $90\text{ns}$, where the value of $\Delta y$  saturates.
    (c) The topological charge $Q$.
    }
    \label{fig6}
\end{figure}

\begin{figure}[h]
    \centering
    \includegraphics[width=\columnwidth]{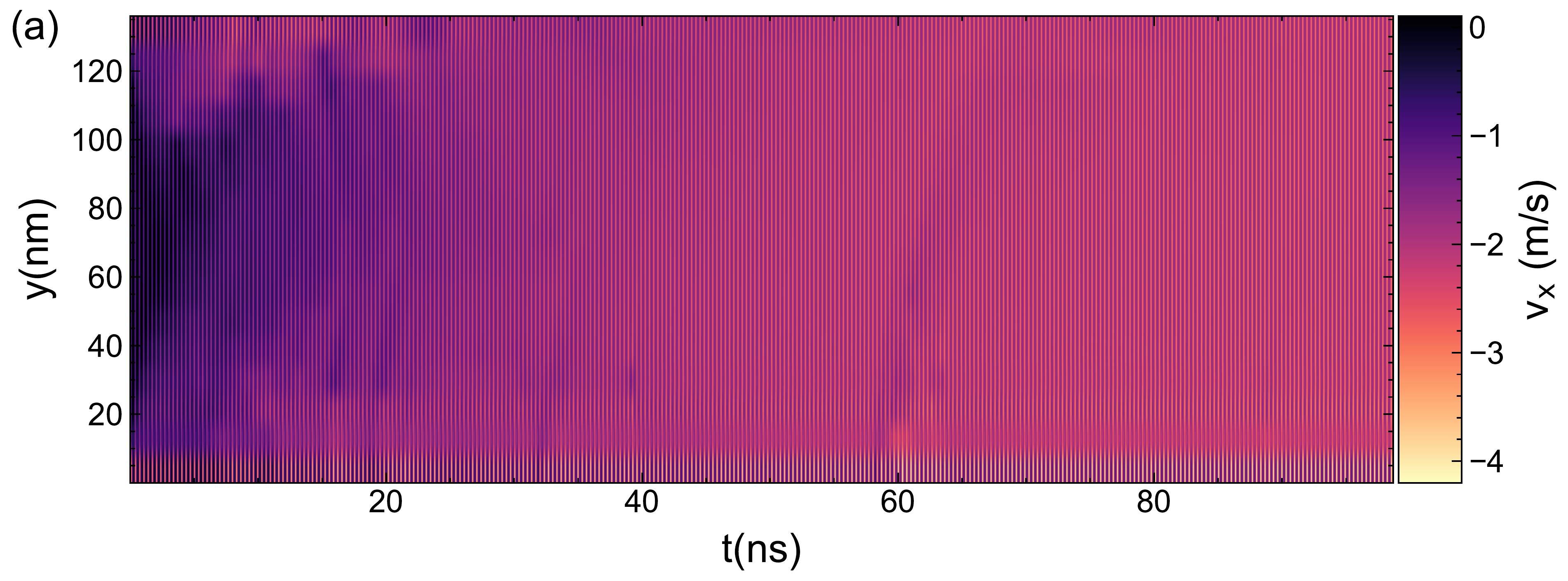}
    \includegraphics[width=\columnwidth]{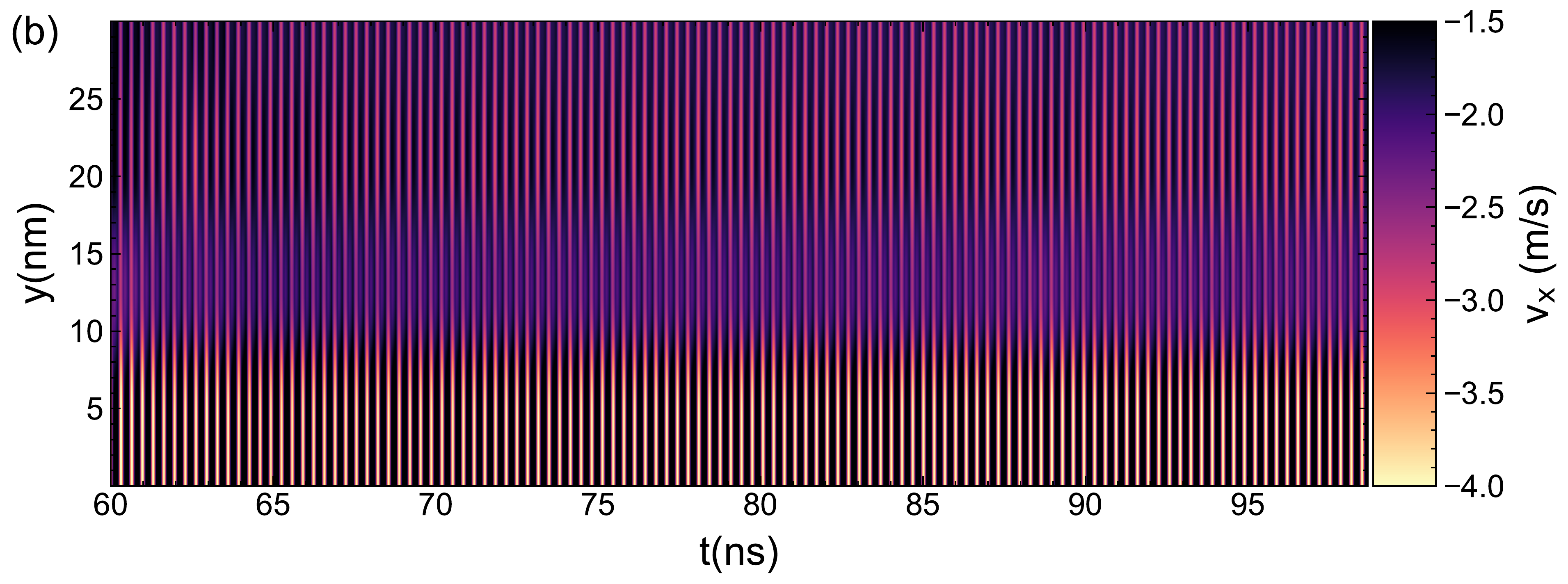}
    \caption{(a) Heatmap of skyrmion velocity $v_x$
      as a function of the distance from the wall $y$ and
      the time $t$ for rectified ac driving with
      $j_0=8.01\times10^9\text{A/m}^2$. Lighter colors
      indicate more rapidly moving skyrmions.
      (b) A blow up of panel (a) showing
      the velocities near the wall for $60<t<100$ ns to
      more clearly illustrate the shear banding
	effect.}
    \label{fig7}
\end{figure}

In figure~\ref{fig6}(a) we plot the skyrmion average velocity per
ac drive cycle 
$\left\langle v_x \right\rangle_c$ and $\left\langle v_y \right\rangle_c$ as a function of time, $t$.
Here we average
the velocity over all skyrmions during the duration of a given ac cycle.
The general behavior of the velocity curves is similar to that found under dc driving. The main 
difference is in the magnitude of the velocities, where we find
that ac driving produces slower motion along the $-x$ direction
compared to dc driving.
There is also a reduction in the length of time during which the magnitude of
the perpendicular velocity is smaller than that of the parallel
velocity, and we find that
$|\left\langle v_x \right\rangle_c| > |\left\langle v_y \right\rangle_c|$
after $t=1.15$ ns.
We obtain saturation velocity values
of 
$\langle v_x \rangle_c = -2.2$ m/s and $\langle v_y \rangle_c = 0$ m/s.
In figure~\ref{fig6}(b) we plot the skyrmion displacements
$\Delta x$ and $\Delta y$
from the $t=0$ positions as a function of time.
The maximum displacement in the
$y$ direction due to compression is $\Delta y^{max}=-17.6$ nm,
smaller than the value found for dc driving.
Figure~\ref{fig6}(c) shows the average topological charge
during the ac cycle, $\langle Q \rangle$, 
as a function of time, $t$.
The magnitude of $|\langle Q \rangle|$ 
decreases with time, similar to what we observe for dc driving.

The main difference between ac and dc driving is visible
in figure~\ref{fig7}, where we plot
a heatmap of the skyrmion velocities $v_x$
as a function of distance from the wall $y$ and time $t$.
This plot provides detailed
information on the distribution of the skyrmion velocities inside the 
sample.
For $t<20$ ns, the skyrmions in the upper and lower parts of the sample are
moving the most rapidly while the
skyrmions in the central part of the sample remain mostly stationary.
This is similar to the observations
made by Zhang {\it et al.} \cite{zhang_structural_2022}
and to what we find under dc driving.
What is different is that
the skyrmions very close to the wall, $y<10$ nm, exhibit much larger
magnitudes of $v_x$
than skyrmions elsewhere in the sample.
This is emphasized in figure~\ref{fig7}(b),
where we plot a blowup of the heatmap in the region near the wall
for $60<t<100$ ns. 
The compression cycle of the ac drive increases the pressure on the
skyrmions at the bottom of the sample,
leading to
deformations in the
skyrmion lattice structure and a reduction in the size of the skyrmions.
During the relaxation cycle, these deformed skyrmions relieve the
pressure by expanding back out to their equilibrium sizes
and pushing the rest of the skyrmions
away from the wall.
The repeated cycle of compression and release
acts as a skyrmion flux pump
that accelerates
the skyrmions in the bottom part of the sample.
This process
also produces a shear
banding effect in which the
skyrmions closest to the wall move much
more rapidly than the other skyrmions.
Skyrmion velocities in the region $y<10$ nm can reach values of up
to $v_x = 4.2$ m/s, while the average velocities elsewhere in the
sample are
around $v_x = 3.0$ m/s.

\section{The influence of $j_0$}

\begin{figure}[h]
    \centering
    \includegraphics[width=0.6\columnwidth]{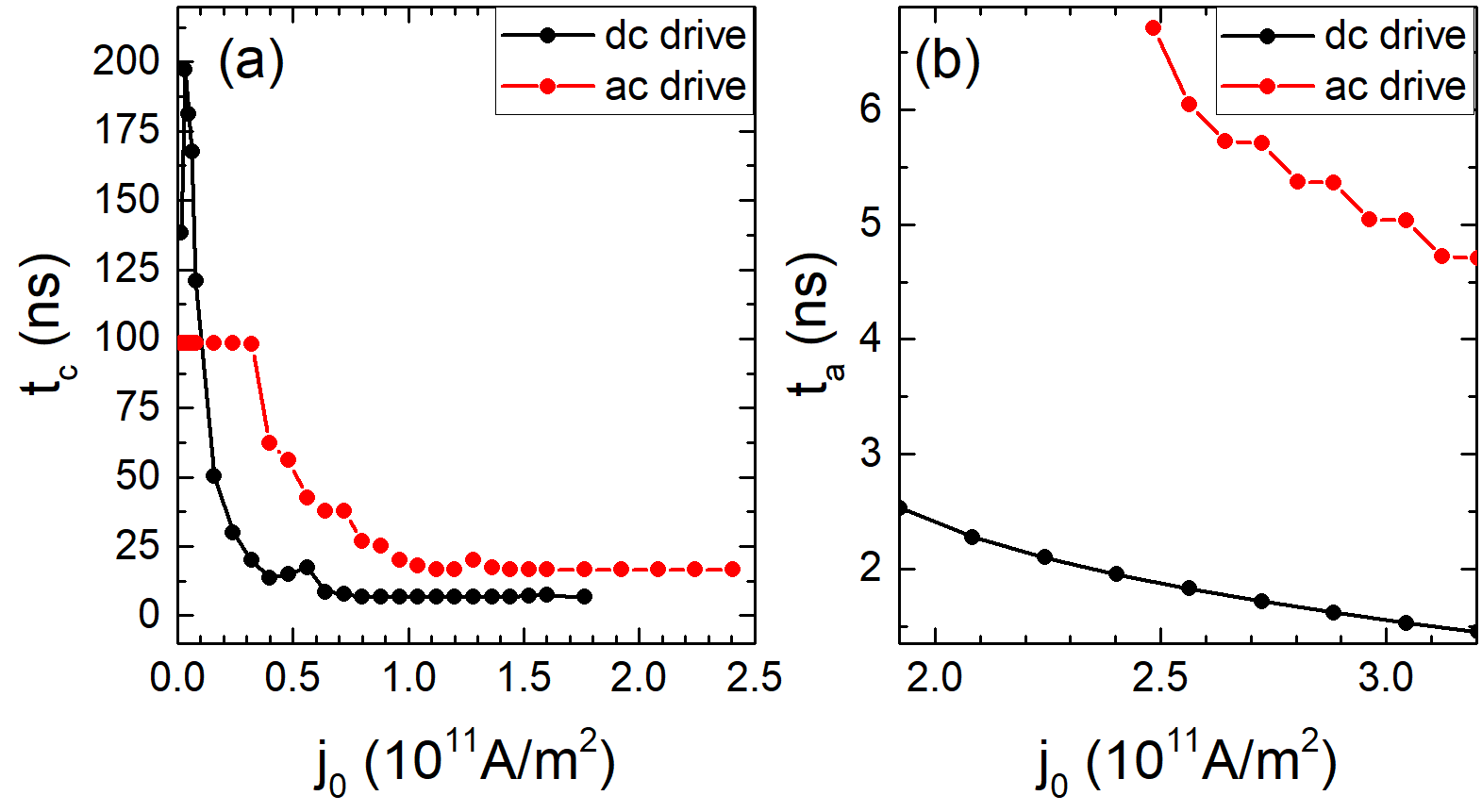}
    \includegraphics[width=0.6\columnwidth]{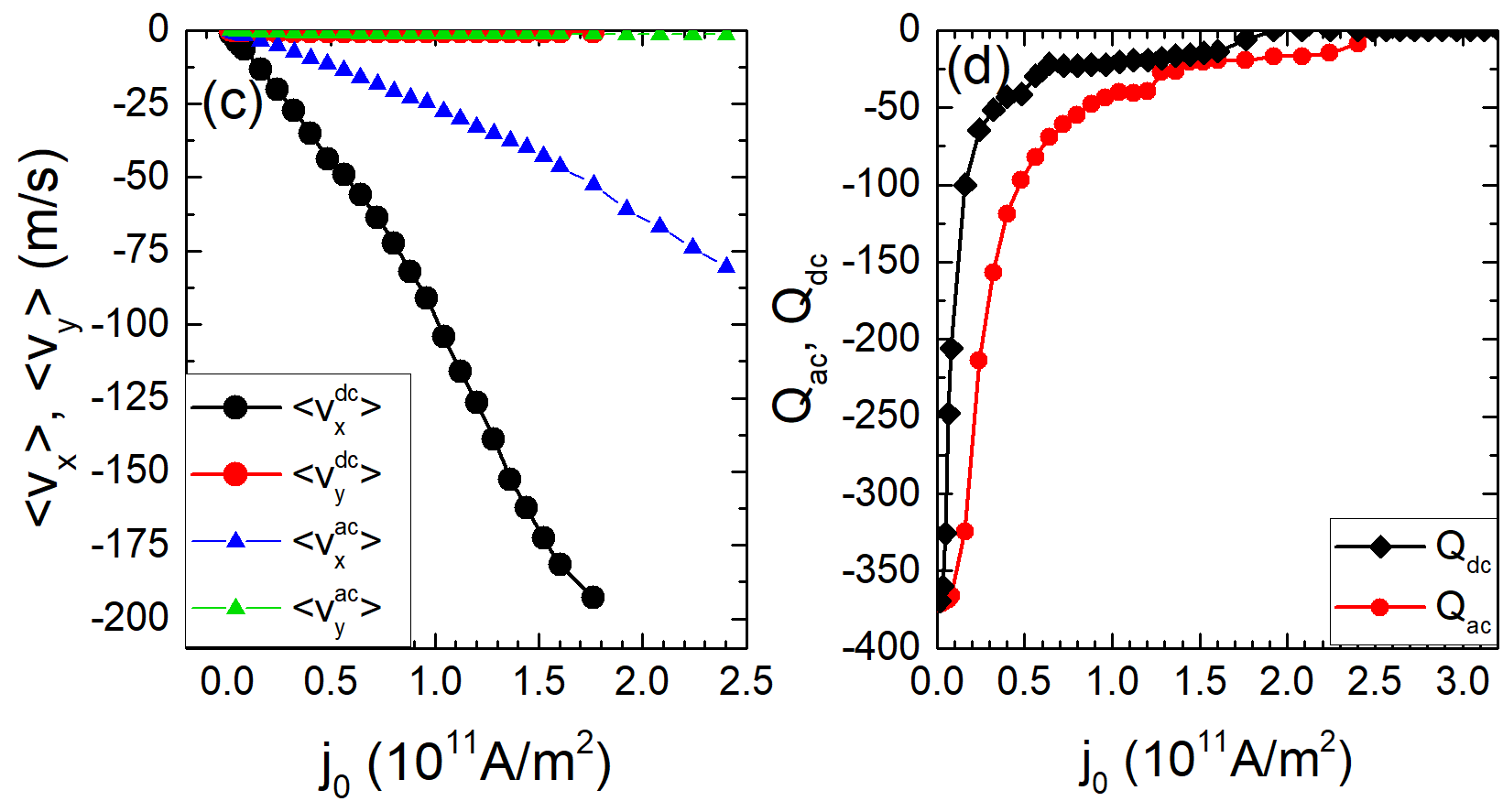}
    \caption{
    Comparison of dc and ac driven systems at
    varied values of $j_0$.
    (a) The time $t_c$ at which the skyrmion lattice reaches a fully
    compressed state vs $j_0$.
    (b) Annihilation time $t_{a}$ vs $j_0$.
    (c) Skyrmion velocities averaged over all skyrmions during the entire
    simulation,
    $\langle v_x \rangle$ and $\langle v_y \rangle$, vs $j_0$ for
    ac and dc driving.
    (d) The topological charge $Q_{dc}$ and $Q_{ac}$ vs $j_0$.
    }
    \label{fig8}
\end{figure}

We  next analyze how varying the
magnitude of the external drive affects the skyrmion lattice
structure and annihilation processes.
Figure~\ref{fig8} shows results for both dc and ac driving.
In figure~\ref{fig8}(a)
we plot the time $t_c$ at which the skyrmion
lattice reaches its maximum compression
versus $j_0$. For $t>t_{c}$, the
skyrmion adopts a stable configuration.
For low drives of
$j_0<1.6\times10^{10}\text{A/m}^2$,
$t_{c}$ is smaller for ac driving than for dc driving.
This is because the
shear banding compression and relaxation cycles
in the ac driven system more rapidly compactify the skyrmion lattice.
In contrast, for $j_0>1.6\times10^{10}\text{A/m}^2$,
the skyrmion lattice compresses faster under dc driving
since the constant pushing force at these higher currents
becomes more efficient than the shear banding organization process.

We find that there is a critical current $j_0^{crit}$ above which
the annihilation of skyrmions does not stop but 
continues until all of the skyrmions have been annihilated at the wall.
We obtain 
$j_0^{crit} = 1.76\times 10^{11}\text{A/m}^2$ for dc driving
and $j_0^{crit} = 2.4\times 10^{11}\text{A/m}$
for ac driving.
For $j_0>j_0^{crit}$,
it still takes a time $t_a$ for all of the skyrmions to annihilate completely.
In figure~\ref{fig8}(b)
we plot $t_{a}$ versus $j_0$ for dc and ac driving.
The skyrmion lattice survives almost three times as long under ac
driving compared to dc driving.
In the case of ac driving, the skyrmions are being pushed towards the wall
during the compression cycle, but
during the relaxation cycle the skyrmions have the same amount of
time to relieve the pressure, reorganize their lattice, and
avoid annihilation.
In contrast, the dc drive system constantly pushes
the skyrmions towards the wall, continuously shrinking the skyrmions 
and favoring their annihilation.

In figure~\ref{fig8}(c) we plot the skyrmion velocities $\langle v_x\rangle$
and $\langle v_y\rangle$ averaged over all skyrmions during the entire
simulation as a function of $j_0$ for dc and ac driving
up to the critical annihilation currents.
For both types of driving,
$\langle v_y \rangle \approx 0$ since the skyrmions are
simply being compressed in the $y$ direction,
while $\langle v_x \rangle$
increases monotonically in magnitude with
increasing $j_0$.
The dc driven
system exhibits much
larger values of $|\langle v_x \rangle|$ than the ac driven system. 

\begin{figure}[h]
    \centering
    \includegraphics[width=0.8\columnwidth]{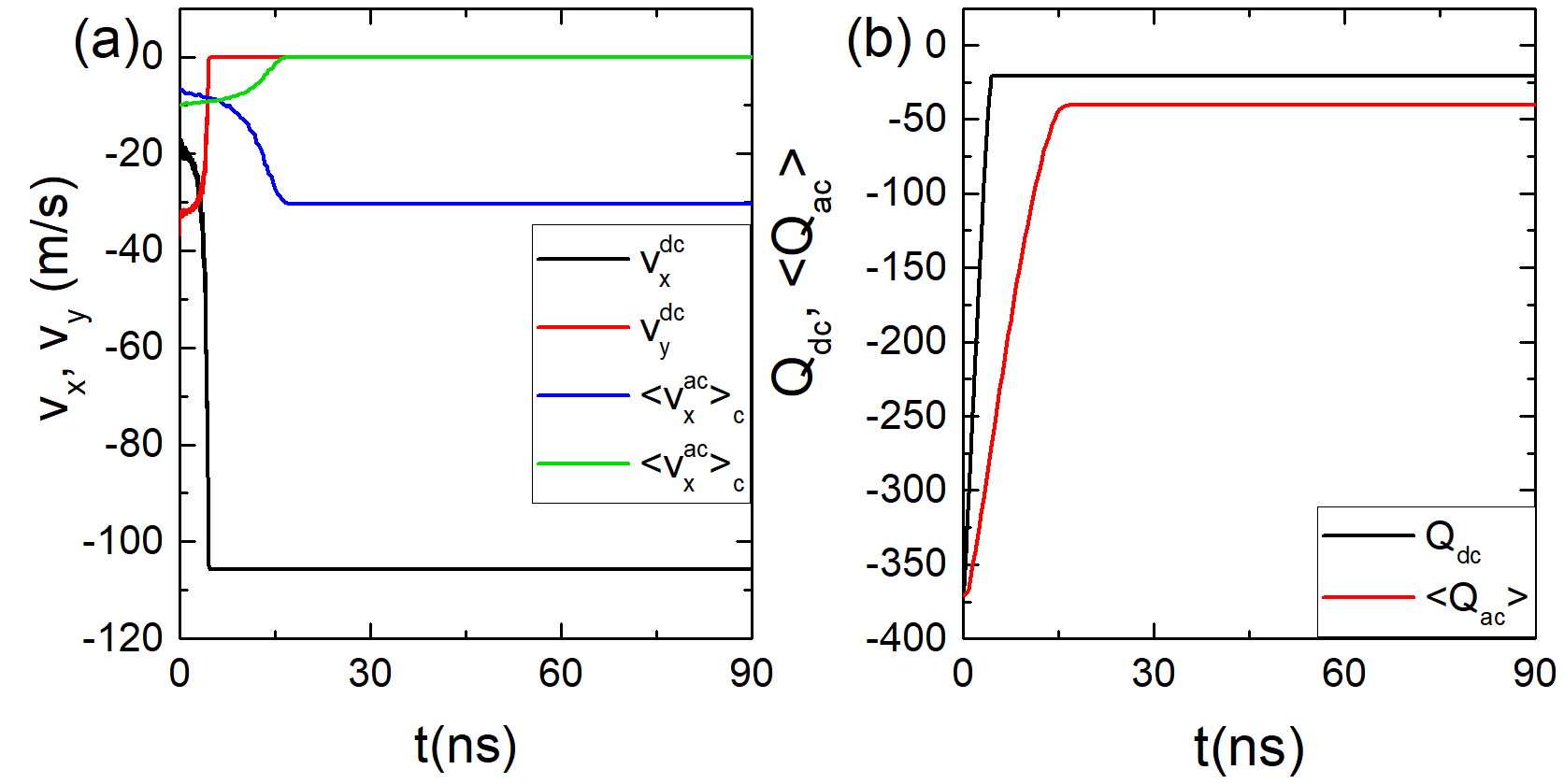}
    \includegraphics[width=0.8\columnwidth]{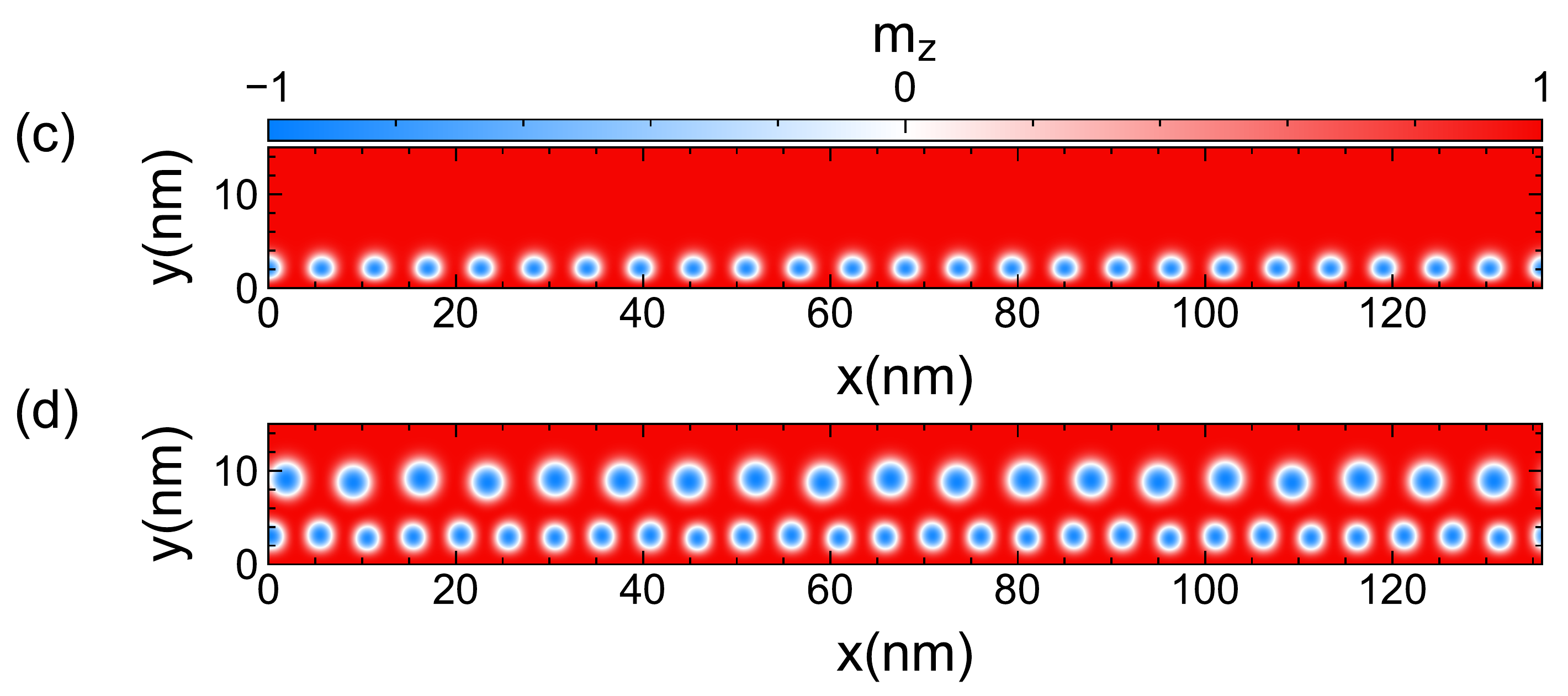}
    \caption{Comparison of dc and ac driven systems at
      $j_0=1.04\times10^{11}\text{A/m}^2$.
      (a) Skyrmion instantaneous (for dc driving)
      or cycle-averaged (for ac driving) velocities vs time.
      The dc driven system exhibits much 
    faster skyrmion transport along the $-x$ direction.
    (b) Topological charge $Q$ vs time. The ac driven
    system preserves a larger magnitude of topological charge.
    (c,d) Images of the spin configurations at $t=90$ ns.
    (c) In the dc driven system, one row of skyrmions is
    stabilized, while (d) in the ac driven system, two skyrmion rows
    are stabilized.
    }
    \label{fig9}
\end{figure}

Figure~\ref{fig8}(d) shows the charges $Q_{ac}$ and $Q_{dc}$ at the stable time 
$t_{c}$ for ac and dc driving as a function of $j_0$.
For every value of $j_0$, $|Q_{ac}| > |Q_{dc}|$,
meaning that for a given $j_0$, the dc driven system 
always contains fewer and/or smaller skyrmions than the ac
driven system. In addition, the annihilation process occurs faster
under dc driving than under ac driving.
There is a wider range of $j_0$ values where the skyrmions remain
stable in the ac driven system;
however, this stability comes with a
velocity cost, since
the skyrmions subjected to ac driving have lower values
of $|\langle v_x\rangle|$.
For a better visualization of the stability, in
figure~\ref{fig9} we compare ac and dc driven systems with
$j_0=1.04\times10^{11}\text{A/m}^2$.
Figure~\ref{fig9}(a) shows the skyrmion velocities as a function of time
and in figure~\ref{fig9}(b) we plot
the topological charge as a function of time.
The ac driven system requires $t_c\sim20\text{ns}$
to reach a stable compression, while for
the dc driven system stable compression is achieved
in $t_c\sim8\text{ns}$.
Skyrmion annihilation occurs for $t<t_c$ in both the ac and dc driven
systems;
however,
the ac driven system has a higher topological charge, or higher skyrmion
density, in the stable compressed state.
This is illustrated in figure~\ref{fig9}(c,d), where the
ac driven system contains two rows of skyrmions moving along
the wall, while there is only one row of skyrmions along the wall
in the dc driven system.
Figure~\ref{fig9}(a) shows that even though the ac driven
system conserves a larger number of skyrmions,
the average velocity is much reduced compared to the dc system.
Skyrmion annihilation occurs as a result of the applied current
pushing the skyrmions against the wall, combined with the
force gradient from the skyrmions in the upper part of the sample, 
which have larger sizes due to the density 
gradient. 
As more and more of the skyrmions are annihilated, 
the pressure from skyrmions in the upper part of the sample vanishes
and eventually
only the spin current pressure
is available for annihilating the last row of skyrmions. 

In possible applications where skyrmions carry information, the difference
in the response between ac and dc driving is important,
since although a dc drive transports
the skyrmions more rapidly, the annihilation risk 
is higher.
The optimum setting would be to use ac driving with
$j_0 < 2.4\times10^{11}\text{A/m}$; however,
variation of other parameters such 
as ac frequency should also be considered.

\section{Varying the skyrmion size with $D/J$}

We next consider the 
influence on the dynamics
of the skyrmion size,
which is controlled by the value of $D/J$. 
In order to avoid annihilation and focus on the dynamics, we fix
$j_0 = 3.2\times10^{9}\text{A/m}^2$ for the dc driven system
and $j_0=8.01\times10^9\text{A/m}^2$ for the ac
driven system.

\begin{figure}[h]
    \centering
    \includegraphics[width=0.32\columnwidth]{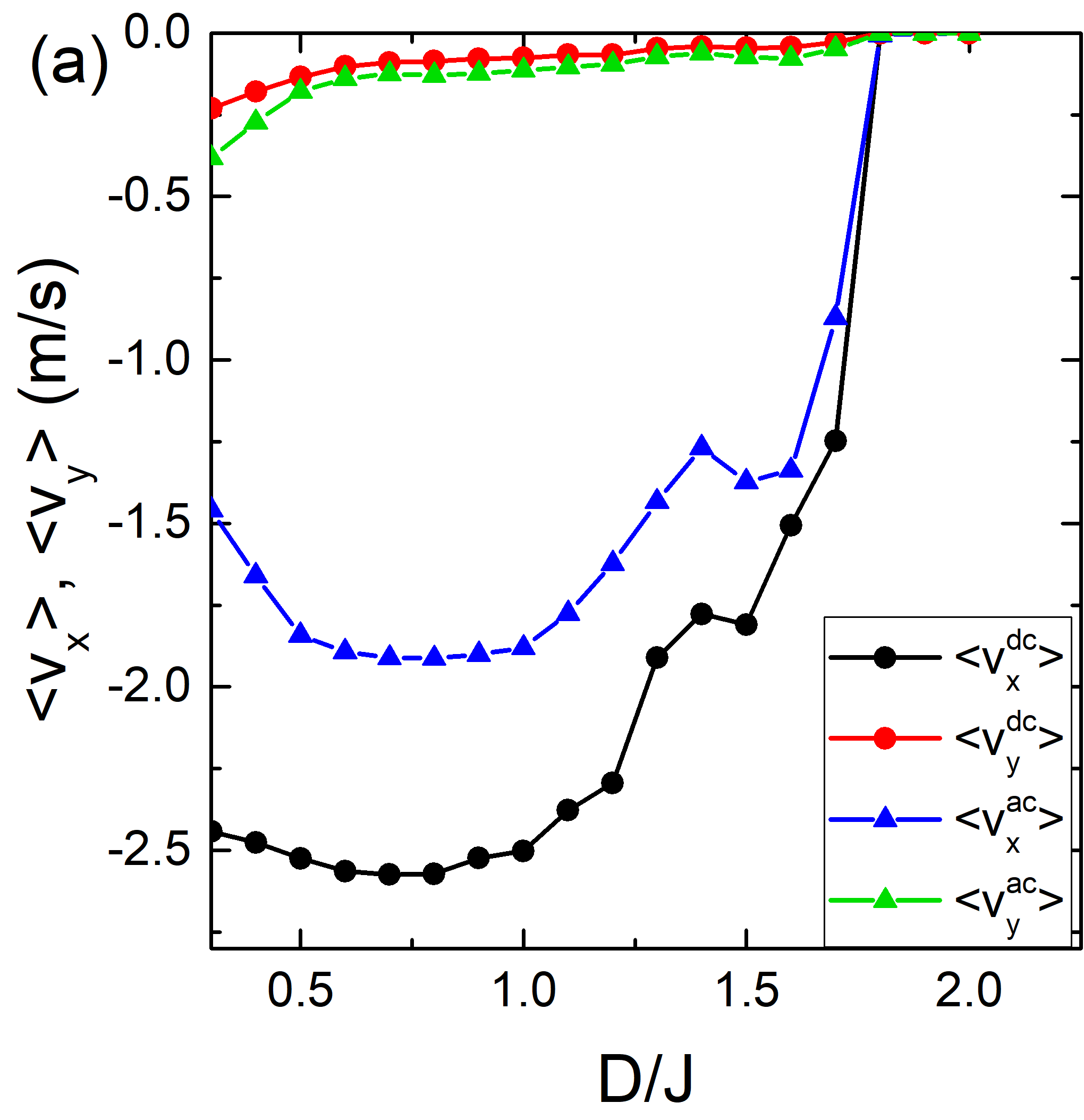}
    \includegraphics[width=0.4\columnwidth]{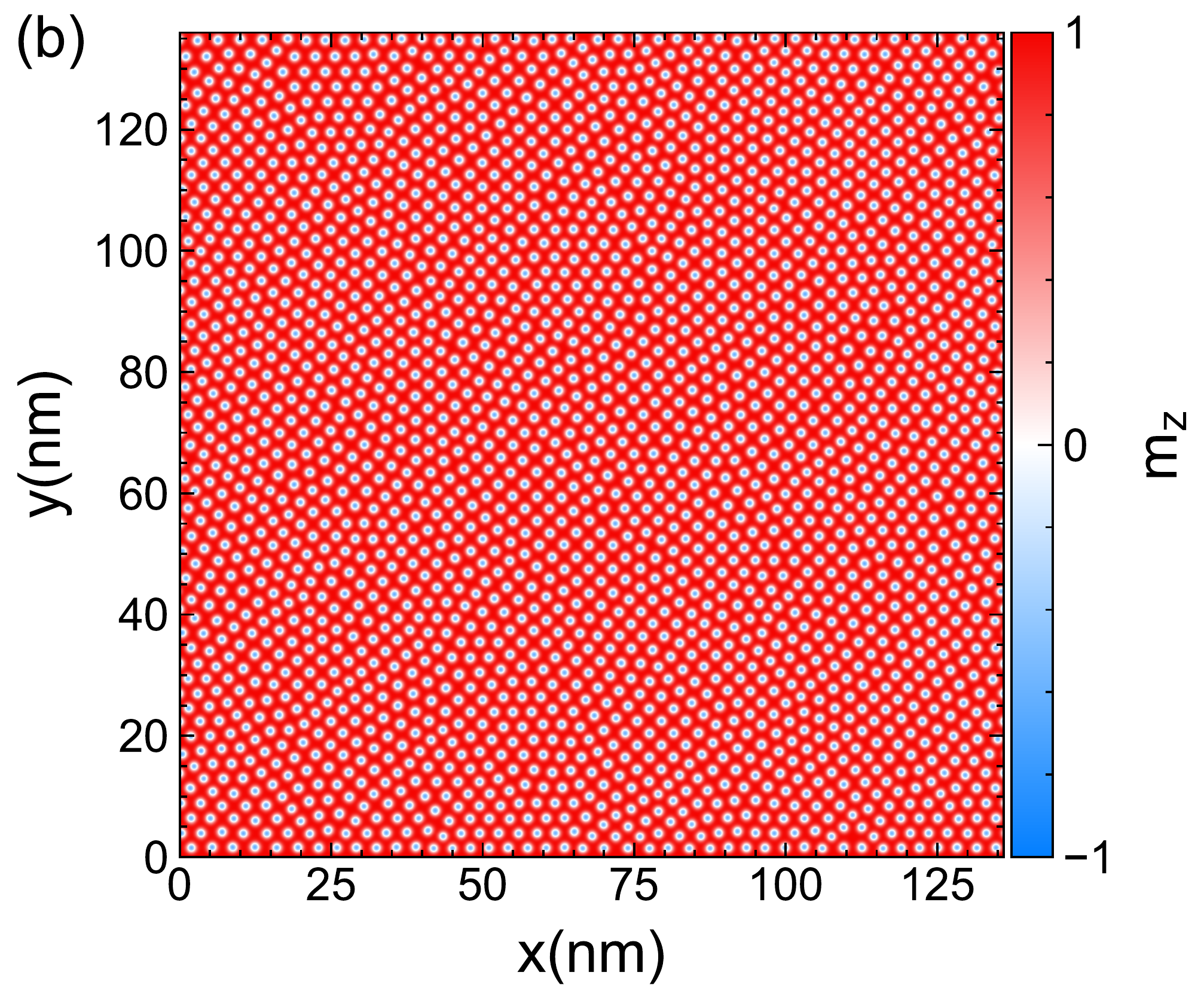}
    \caption{(a) Velocities averaged over the entire simulation in the $x$
      and $y$ directions
      vs $D/J$ in samples with dc driving at
      $j_0=3.2\times10^{9}\text{A/m}^2$ or ac driving at
    $j_0=8.01\times10^9\text{A/m}^2$.
      For both types of driving,
      when $D/J=1.8$
      there is a drop in the magnitude of the $x$ velocity from a
      finite value to zero.
      (c) Spin configuration for a
      dc driven system with $H = 0.5D^{2}/J$, $D/J=1.8$, and $j_0=3.2\times10^{9}\text{A/m}^2$.
    }
    \label{fig10}
\end{figure}

In figure~\ref{fig10} we show the time averaged skyrmion velocities
$\left\langle v_x\right\rangle$ and
$\left\langle v_y\right\rangle$ as a function of $D/J$ for both ac and dc
driving.
There is a clear nonmonotonic behavior of the velocities in the $x$
direction.
When $D/J<0.7$, for both types of driving
the magnitude of the $x$ velocity increases with increasing $D/J$ while
the magnitude of the $y$ velocity decreases.
The largest $x$ direction velocity magnitude appears
around $D/J=0.7$, where
$|\langle v_{x}^{dc}\rangle|\approx 2.6\text{m/s}$ and 
$|\langle v_{x}^{ac}\rangle|\approx 1.9\text{m/s}$.
The magnitude of the $x$ velocity decreases with increasing $D/J$ above
$D/J> 0.7$,
until for $D/J=1.8$
the motion in both the $x$ and $y$ directions
for both ac and dc driving vanishes.
When $D/J$ increases, the skyrmion size decreases, and 
since we are applying a constant field of $H = 0.5D^{2}/J$, 
the skyrmion density increases as the skyrmion size is reduced. 
As the skyrmion gets smaller, it also becomes less susceptible to deformations
and the gradient in skyrmion size from the top to the bottom of the sample
diminishes.
The skyrmion lattice itself also becomes 
more rigid as the skyrmions shrink,
reducing its mobility under this compression geometry
and 
resulting in
the formation of a static compact lattice,
as shown in figure~\ref{fig10}(b). Comparing
the skyrmions in figure~\ref{fig10}(b)
with those in figures~\ref{fig2} and \ref{fig5}, it is clear that the
skyrmion size is much more uniform for the small skyrmions
at $D/J=1.8$.


\section{The influence of the applied magnetic field $H$} 

In this section we investigate how the applied magnetic field $H$ can influence the
dynamics of the system by modifying
the skyrmion density,
size, and stability.
For weaker fields the skyrmions are larger, while for higher fields the 
skyrmions become smaller and
are reduced in number due to the
strengthening of the ferromagnetic background.
We fix $D/J=0.5$ and set
$j_0=3.2\times10^{9}\text{A/m}^2$ for dc driving
and $j_0=8.01\times10^9\text{A/m}^2$ for ac driving.
We do not consider fields $H<0.42 D^2/J$, since
at these smaller fields, other topological textures become intermixed
with the skyrmions during the initialization of the system.

\begin{figure}[h]
    \centering
    \includegraphics[width=0.4\columnwidth]{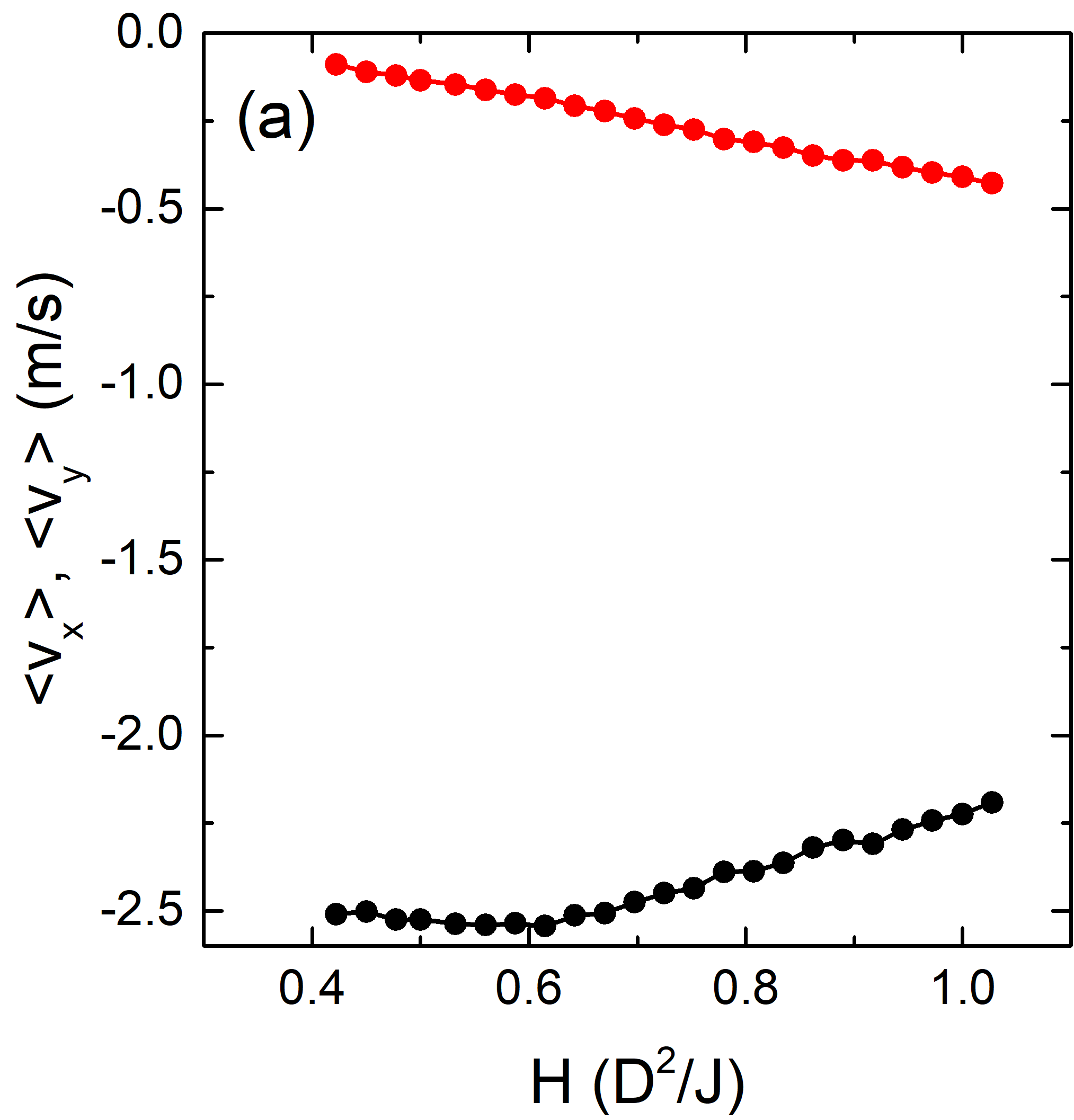}
    \includegraphics[width=0.4\columnwidth]{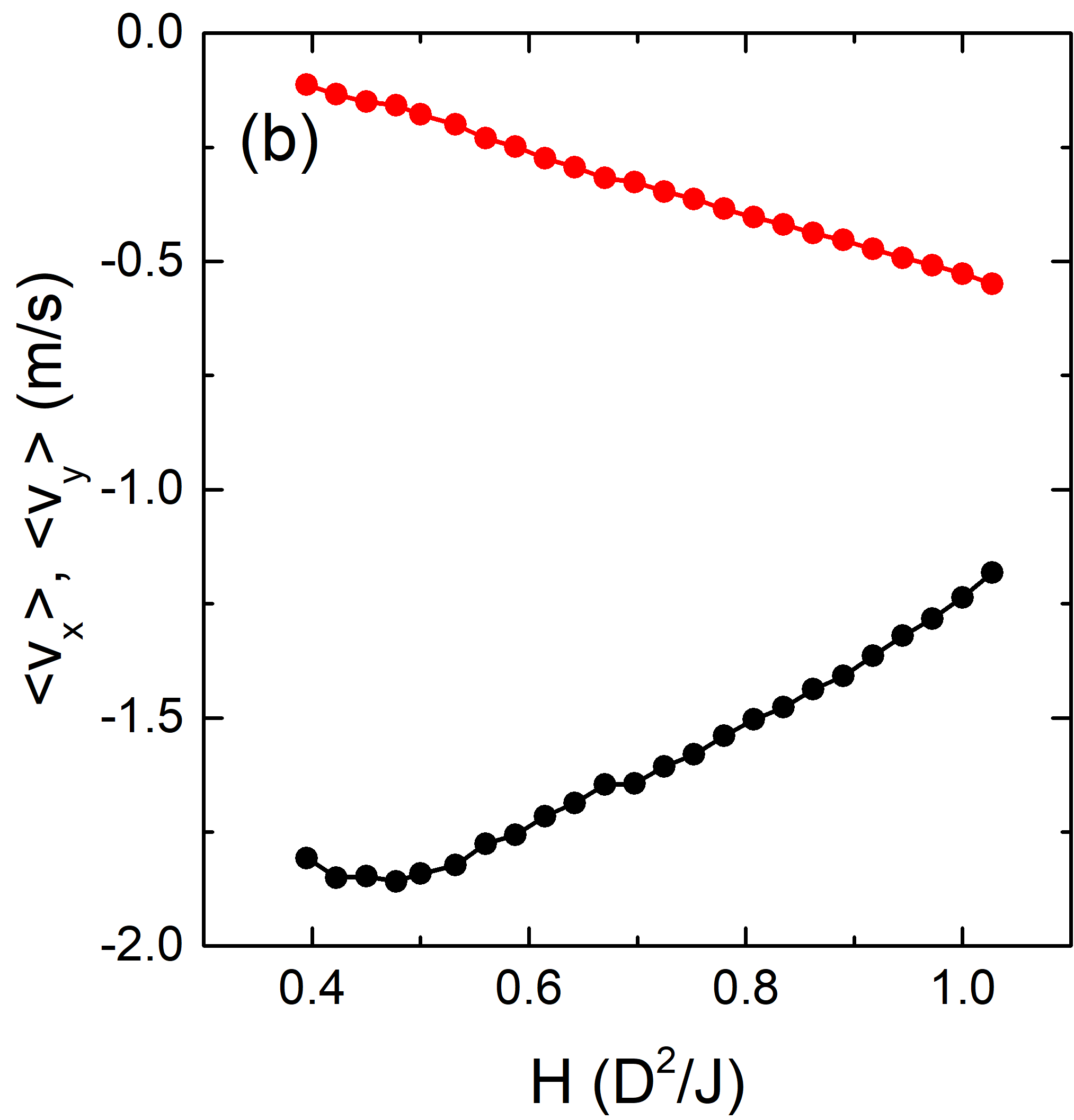}
\caption{Plot of the skyrmion velocity averaged over the entire simulation,
$\left\langle v_x\right\rangle$ (black)
and $\left\langle v_y\right\rangle$ (red), versus magnetic field $H$.
(a) ac driving with $j_0=8.01\times10^9\text{A/m}^2$.
(b) dc driving with $j_0=8.01\times10^9\text{A/m}^2$.
As $H$ increases,
the magnitude of $\left\langle v_x\right\rangle$ decreases
and the magnitude of $\left\langle v_y\right\rangle$
increases for both ac and dc driving.
The decrease in $|\left\langle v_x\right\rangle|$
is a result of the decreasing skyrmion size,
as shown in figure~\ref{fig12}.
The increase in $|\left\langle v_y\right\rangle|$
is also a result of the decreasing
skyrmion size, since as the skyrmions become smaller,
there is more space for compression,
resulting in a higher velocity towards the wall.}
    \label{fig11}
\end{figure}

In figure~\ref{fig11} we plot the time averaged skyrmion velocities
 $\left\langle v_x\right\rangle$ and $\left\langle v_y\right\rangle$
for both dc and ac driving.
In each case, we find a linear increase in
$|\left\langle v_y\right\rangle|$
with increasing $H$ due to the decrease in the size of the skyrmions. 
At high values of $H$, the skyrmions become smaller
but their density does not vary 
appreciably. As a result, the reduced size of
the skyrmions produces a more dilute skyrmion lattice
in which the skyrmions can more freely move.
The vacant regions
produce an increase in 
$|\left\langle v_y\right\rangle|$
since the skyrmions must travel a greater distance
in the $y$ direction to compress the lattice.
The behavior of $\left\langle v_x\right\rangle$ is non-monotonic
as $H$ varies, and there is a value of magnetic field at which
$|\left\langle v_x\right\rangle|$ reaches its
greatest magnitude.
For dc driving, the $x$ velocity begins to decrease in
magnitude for $H>0.6D^{2}/J$,
while for ac driving, this decrease
occurs for $H>0.5D^{2}/J$.

\begin{figure}[h]
    \centering
    \includegraphics[width=0.3\columnwidth]{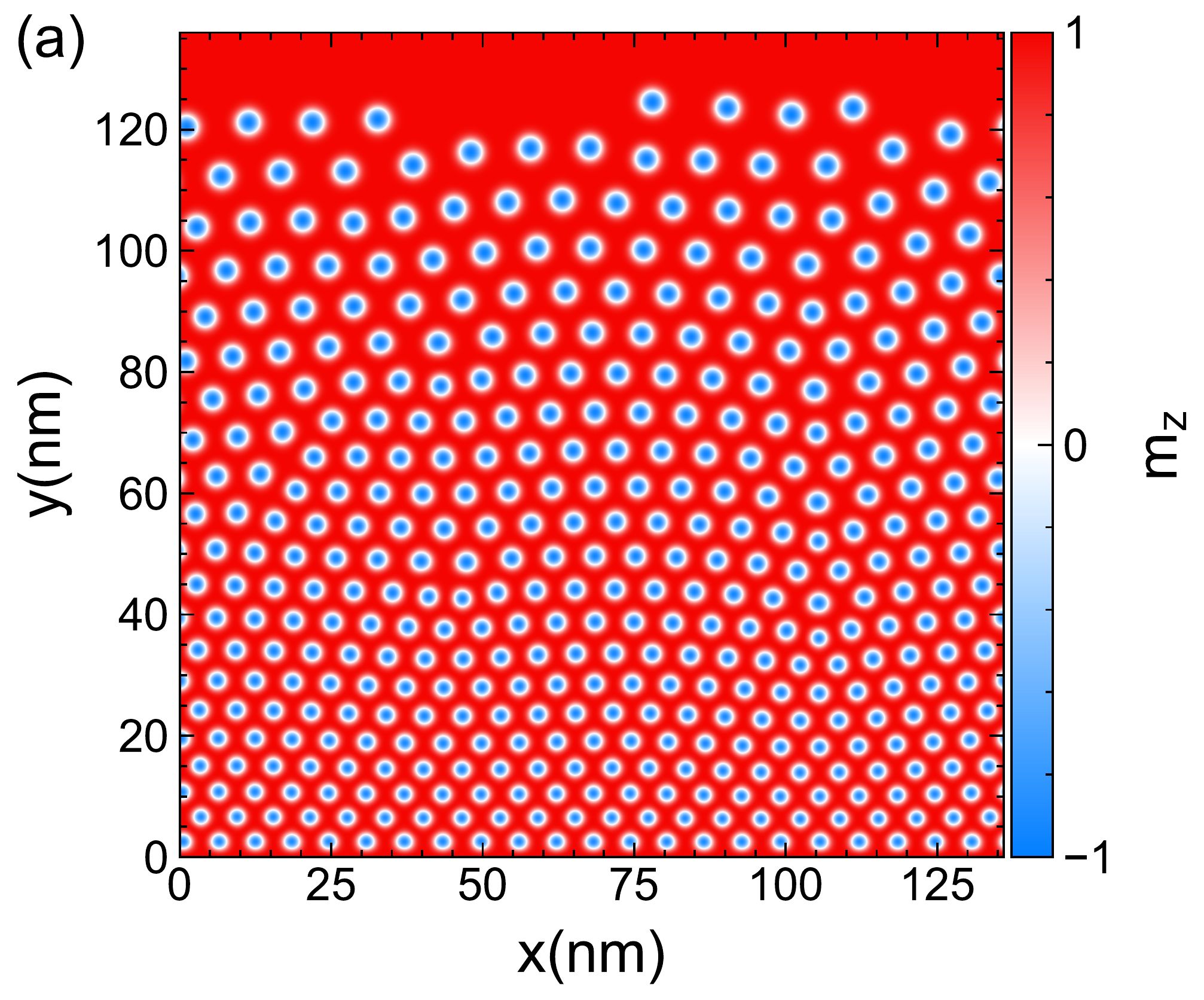}
    \includegraphics[width=0.3\columnwidth]{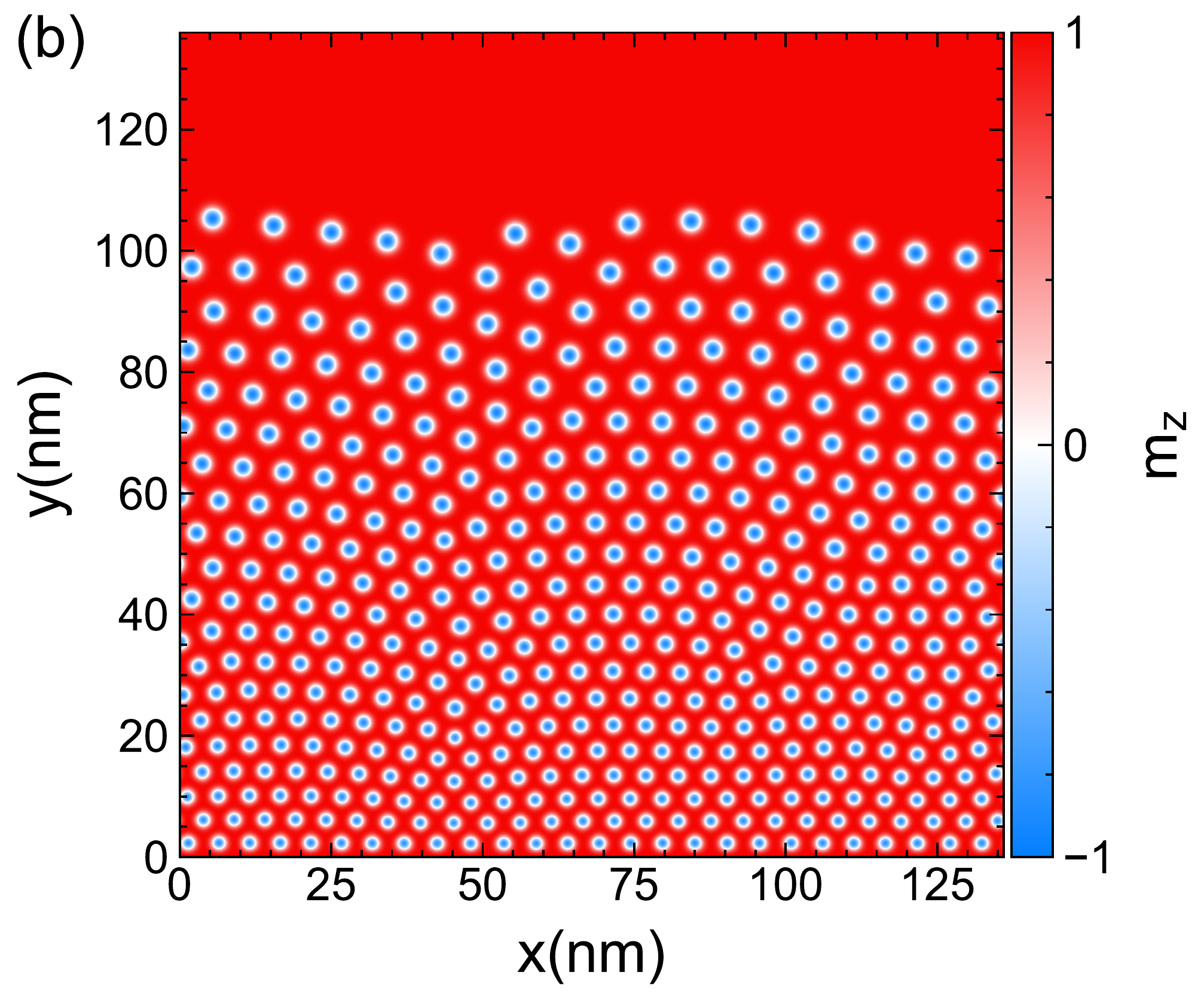}
    \includegraphics[width=0.3\columnwidth]{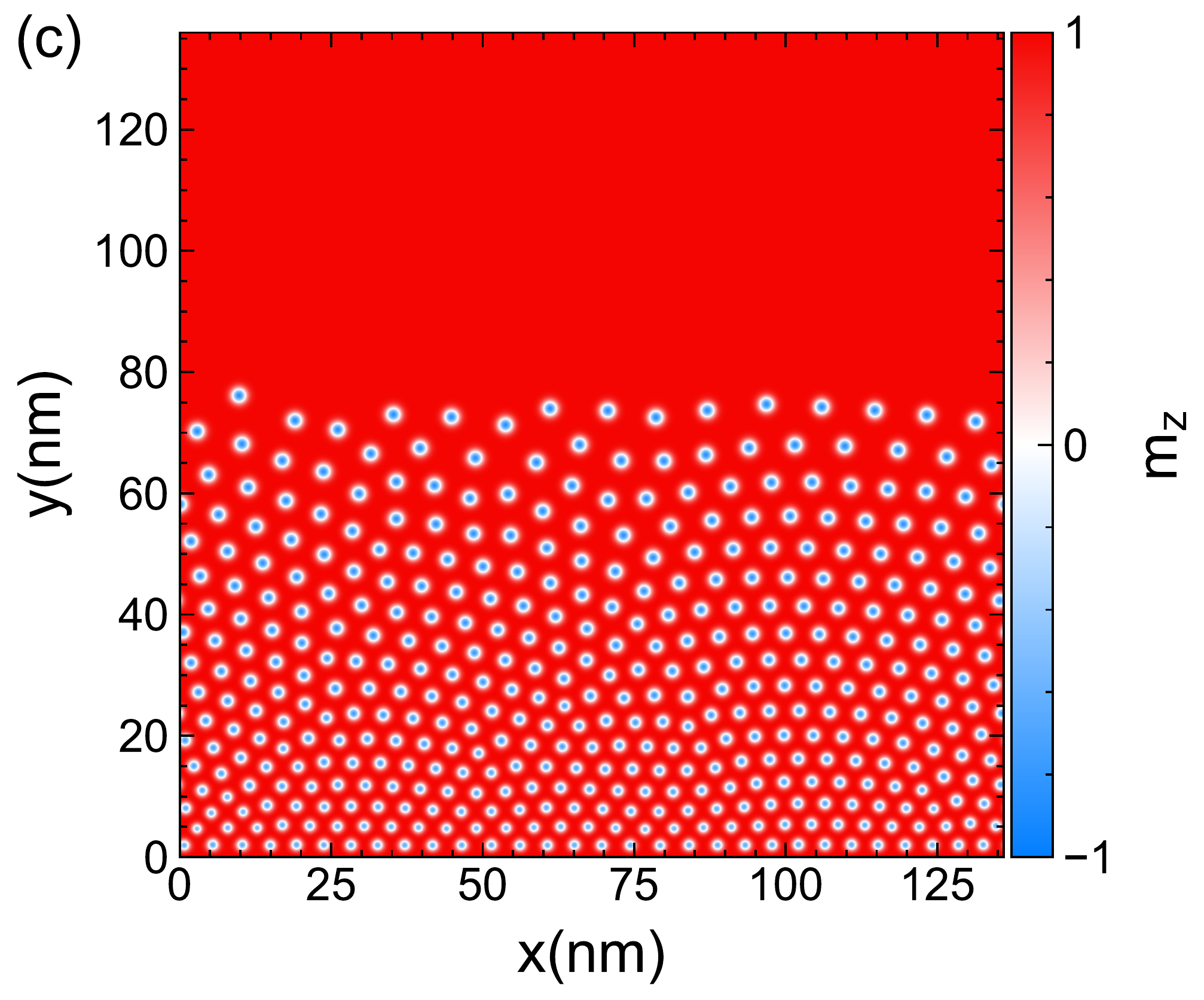}
    \caption{Images of the spin configurations in samples with ac driving
at the end of the simulation      
for systems with (a) $H=0.5325D^2/J$, (b) $H=0.6425D^2/J$,
and (c) $H=0.89D^2/J$.
As the magnetic field increases,
the skyrmion size decreases, leaving more space for
compression, as indicated by the greater displacement of the lattice
in the $y$ direction with increasing $H$.
Smaller skyrmions are more rigid and
suffer fewer deformations,
making it more difficult for the lattice to translate
along the $x$ direction and resulting in a decreased velocity
parallel to the wall.}
    \label{fig12}
\end{figure}

An illustration of the skyrmion sizes and the vacant spaces that appear
in the skyrmion lattice under different magnetic fields appears
in figure~\ref{fig12}.
As $H$ increases, the skyrmion size decreases, creating empty space and
enabling a greater compression
of the lattice along the $y$ direction.
In figure~\ref{fig12}(a) at $H = 0.5325D^2/J$, the 
skyrmions are still relatively large and there is a clear gradient in
skyrmion size from top to bottom.
At $H = 0.6425D^2/J$ in figure~\ref{fig12}(b), the skyrmions 
are reduced in size, while for
$H = 0.89D^2/J$ in figure~\ref{fig12}(c), 
the skyrmions have become very small,
have lost their size gradient, and become tightly
packed into the bottom part of the sample.
The skyrmion sizes in figure~\ref{fig12}(c)
are similar to what is shown in figure~\ref{fig10}(b),
and there is no deformation of the skyrmions close to the wall
since the skyrmions are small enough to have become rigid.
Even without deformation of the skyrmions,
we still find motion along the $-x$ direction.
This indicates that motion parallel to the wall results from
a combination of both
the skyrmion size gradient and the skyrmion density
gradient. As more skyrmions are added to the sample, it becomes
more difficult to compress and move the lattice,
destroying the motion along the $-x$ direction.

\section{Particle Based Model}

In order to better understand the role of the changing size of the skyrmions, 
we have also considered a Thiele equation approach
in which the skyrmions are treated as point-like
particles
that have a repulsive interaction with each other
and move under damping and a Magnus term, similar to what has been
employed in previous studies \cite{vizarim_skyrmion_2020,vizarim_skyrmion_2020a,vizarim_topological_2020,vizarim_shapiro_2020,vizarim_soliton_2022,reichhardt_commensuration_2022}.   
The equation of motion for a skyrmion $i$ is

\begin{equation}
	\alpha_{d}{\bf v}_{i} + \alpha_{m} {\bf z}\times {\bf v}_{i} =  
	{\bf F}^{ss}_{i} + {\bf F}^{wall}_{i} +  {\bf F}^{drive}_{i}.
\end{equation}

\begin{figure}
    \centering
    \includegraphics[width=0.5\columnwidth]{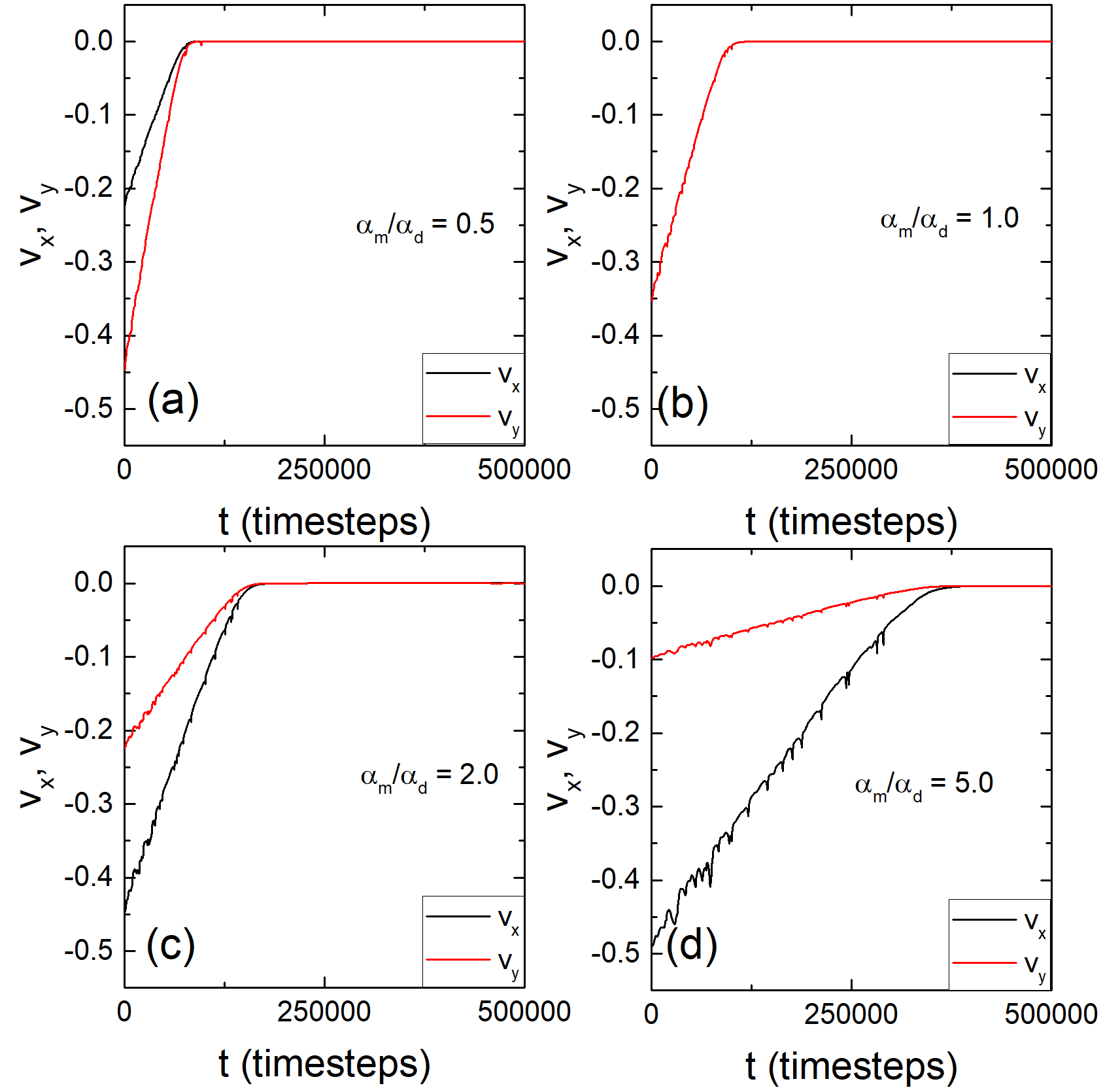}
\caption{Response as a function of time in a particle-based model
with dc driving       
for varied ratios of the Magnus to dissipative
terms,  $\alpha_{m}/\alpha_{d} =$ (a) 0.5, (b) 1.0, (c) 2.0, and
(d) $5.0$.
In this case the skyrmions cannot change size and the motion
is transient.} 
\label{fig13}
\end{figure}

Here $\alpha_d$ is the damping term, $\alpha_m$ is the Magnus term, ${\bf F}^{ss}_{i}$ is the
repulsive skyrmion-skyrmion interaction, ${\bf F}^{wall}$ is the interaction with the wall,
and ${\bf F}^{drive}$ is the external driving force. The skyrmion-wall
interaction is repulsive and is given by
$U(r) = U_{0}\exp{(r/a_{0})}^2$. The
skyrmion-skyrmion interaction is
also repulsive and has the Bessel shape 
${\bf F}^{ss}_{i}  =  \sum^{N_{k}}_{i}K_{1}(r_{ij}) {\bf \hat{r}}_{ij}$, 
where $K_{1}$ is the first order Bessel function. 
In this particle based model, there is no skyrmion annihilation
and the skyrmions have no size variations or internal degrees of freedom.
In
figure~\ref{fig13}
we plot the skyrmion velocities
$v_x$
and
$v_y$
versus time for a system containing
100 point-like skyrmions interacting with a wall under a dc drive at
Magnus force to damping force ratios of
$\alpha_{m}/\alpha_{d} = 0.5, 1.0, 2.0$ and $5.0$.   
In each case,
the system exhibits transient skyrmion
velocities that vanish over time.
In the atomistic magnetic model limit, 
for high values of $D/J > 1.7$ the skyrmions are small enough so that 
they also act like point-like particles,
and the skyrmion velocities also drop
rapidly to zero with time. Hence,
the smaller skyrmions of the atomistic model
can be sufficiently well represented using
a particle-based skyrmion model.

\section{Summary}

We have investigated the dynamics and the formation of conformal crystals
for skyrmion crystals 
being pushed against a wall under
dc and rectified ac driving.
For dc driving, the
skyrmion lattice does not compress uniformly
but forms a conformal crystal structure with a 
density gradient and characteristic arch-like structure similar to
that found experimentally in 
magnetic particles under gravity and vortices in type-II superconductors
being injected into a sample from one side.
The conformal crystal is not perfect but
contains topological and structural
defects
similar to those found for the vortices and magnetic particles.
One difference is that in the
skyrmion case there is a skyrmion size gradient, with smaller
skyrmions located close to the wall and larger skyrmions located
further from the wall.
We also find that during compression, although skyrmion
annihilation is possible, there
is an increase in the transverse velocity
due to the Magnus force-induced creation of
flows perpendicular to the pressure gradient direction.
Under rectified ac driving with
two states, a relaxation and a compression cycle,
the rate of skyrmion annihilation is reduced
but there is still a net dc motion transverse to the wall as a result of a
skyrmion Magnus ratchet effect. 
Under dc driving, the skyrmions move as a rigid crystal,
while for ac driving we observe a shear banding effect in which 
skyrmions near the wall move twice as fast as skyrmions far
from the wall.
Upon varying the magnitude of the external drive, 
we find a critical spin current $j_{0}^{crit}$
above which all the skyrmions in the sample are annihilated,
but below which
the skyrmions can form a stable crystal
translating along the $-x$ direction.
To modify the skyrmion size we vary
the ratio $D/J$ and find that
transverse flow occurs only for larger skyrmions with
$D/J < 1.7$.
The smaller skyrmions are
more rigid,
destroying the size gradient and consequently the transverse motion.
We compare our results to those obtained in a particle-based model,
and show that the rigid skyrmions of the particle-based model undergo
no transverse motion, in agreement with what we find for small skyrmions in the
atomistic model.
A gradient in skyrmion size is required to produce steady state
transverse motion, which arises as the result of
an effective gradient in the skyrmion Hall angle
since smaller skyrmions have a higher intrinsic Hall angle. 
Our results are relevant for the construction of devices
in which skyrmions interact with magnetic walls, 
domain walls, or other types of repulsive interfaces.

\section*{Acknowledgements}

This work was supported by the US Department of Energy through the Los Alamos National Laboratory. Los
Alamos National Laboratory is operated by Triad National Security, LLC, for the National Nuclear Security
Administration of the U. S. Department of Energy (Contract No. 892333218NCA000001). 
J.C.B.S and N.P.V acknowledge funding from Fundação de Amparo à Pesquisa do Estado
de São Paulo - FAPESP (Grants 2021/04941-0 and 2017/20976-3 respectively).

\section*{References}
\bibliographystyle{iopart-num}
\bibliography{refs}

\end{document}